\documentclass[preprint,11pt,3p,authoryear]{elsarticle}

\usepackage{svg}
\usepackage{url}
\usepackage{amssymb}
\usepackage{amsmath}
\usepackage{gensymb}
\usepackage{booktabs}
\usepackage{multirow}
\usepackage{threeparttable}
\usepackage{graphicx}
\usepackage{caption}
\usepackage{array}
\usepackage{subcaption}
\usepackage{cleveref}
\usepackage{amsthm}
\usepackage{geometry}
\usepackage{tabularray}
\usepackage{setspace}
\usepackage{lineno}

\usepackage[final]{changes}
\definechangesauthor[color=purple, name={Yongjin Choi}]{YC}

\journal{Computers and Geotechnics}

\begin{document}
\doublespacing
\begin{frontmatter}

\title{Inverse analysis of granular flows using differentiable graph neural network simulator}

\author[label1]{Yongjin Choi}
\author[label1]{Krishna Kumar}

\affiliation[label1]{organization={Maseeh Department of Civil Architectural and Environmental Engineering Austin, The University of Texas at Austin},
            city={Austin},
            postcode={78712}, 
            state={TX},
            country={USA}}

\begin{abstract}
Inverse problems in granular flows, such as landslides and debris flows, involve estimating material parameters or boundary conditions based on a target runout profile. Traditional high-fidelity simulators for these inverse problems are computationally expensive, restricting the number of possible simulations. These simulators are also non-differentiable, making efficient gradient-based optimization methods for high-dimensional spaces inapplicable. Machine learning-based surrogate models offer computational efficiency and differentiability. However, they often struggle to generalize beyond their training data due to relying on low-dimensional input-output mappings that fail to capture the complete physics of granular flows. We propose a novel differentiable graph neural network simulator (GNS) that combines reverse mode automatic differentiation of graph neural networks with gradient-based optimization for solving inverse problems. GNS learns the dynamics of granular flow by representing the system as a graph and predicts the evolution of the graph at the next timestep, given the current state. The differentiable GNS demonstrates optimization capabilities beyond the training data. We demonstrate the effectiveness of our method for inverse estimation across single and multi-parameter optimization problems, including evaluating material properties and boundary conditions for a target runout distance and designing baffle locations to limit a landslide runout. Our proposed differentiable GNS framework solves inverse problems with orders of magnitude faster convergence than the conventional gradient-based optimization approach using finite difference.
\end{abstract}


\end{frontmatter}


\section{Introduction}
\label{sec:intro}
In geotechnical engineering, addressing optimization problems, such as the inverse modeling of landslides to infer material properties from target runout distances and design protective structures, is essential.  Traditionally, this inverse optimization process has been approached by iteratively adjusting material properties to closely match observed results, often resulting in oversimplified, integer-valued material parameters. Design problems focus on optimizing geometries, such as the location and geometry of protective structures, to effectively mitigate and control granular flows~\citep{liu2023efficient, babu2008optimum}. Addressing inverse and design optimization challenges is critical to developing robust models for managing landslide risks~\citep{calvello2017role, cuomo2015inverse, Abraham2021rollout_modeling}.

Inverse analysis often requires multiple forward simulations to adjust parameters until they match the desired outcome. Conventional high-fidelity forward simulators, such as discrete element method (DEM)~\citep{staron2005, kermani2015, kumar2017_lbm-dem} and Material Point Method (MPM)~\citep{mast2014, kumar2017_mpm-dem}, are computationally intensive. This limits their practicality for repeated evaluations and constrains the range of parameters that can be effectively analyzed. The lack of comprehensive parametric sweeps results in unrealistic parameters that fit the target runout or sub-optimal designs. Although simplified depth-averaged models are computationally efficient~\citep{hungr2009two, Christen2010ramms, mergili2017ravaflow}, they often fail to capture the full complexity of granular flow dynamics and are still limited to a finite set of parametric simulations. This issue is compounded as the parameter space becomes more complex and multi-dimensional.

Sampling-based methods like grid search \citep{ensor1997gridsearch}, cross-entropy \citep{ho2010cross-entropy}, and Bayesian optimization \citep{frazier2018tutorial_bayesian} often require a significant number of forward simulations, especially as the number of parameters increases. In contrast, gradient-based optimization utilizes derivatives of the objective function to navigate the parameter space efficiently. This approach often leads to faster convergence with fewer simulations but requires the gradient of the objective function. However, traditional simulators are not differentiable. While finite difference methods can approximate the derivatives, they require a smaller increment size and are susceptible to numerical errors.

A popular gradient-based optimization approach is the adjoint method \citep{pires2001adjoint_tsunami, cheylan2019lbm_shape_adjoint}. The adjoint method computes the gradients by introducing an auxiliary adjoint variable $\lambda$ that satisfies an adjoint equation involving the derivatives of the forward problem. By solving this single adjoint equation, we can efficiently compute the gradients of the objective function $f$ for all variables simultaneously without having to compute the derivatives of $f$ numerically. However, the complexity and mathematical challenges inherent in solving adjoint equations, particularly for non-linear equations or problems involving discontinuities like friction, often necessitate alternative approaches.

In this context, Machine Learning (ML)-based surrogate models emerge as a compelling alternative for their efficiency and differentiability. These models create a non-linear functional mapping between influence factors — such as geometry, boundary conditions, and material properties — and the outcomes of granular flows, like runouts~\citep{zeng2021, Ju2022}. Despite their efficiency, the key limitation of ML-based models lies in their low-dimensionality mapping between a finite set of input parameters and output response, which does not fully encapsulate the underlying physics that governs flow behavior. For example, regression-based ML models could learn the relation between the material property and runout of granular columns but fail to generalize to other geometries or material properties. Consequently, these models often face challenges in generalizing beyond their initial training datasets, indicating a need for surrogate models that capture the complex granular flow behavior.

Recent advancements in learned physics simulators offer a promising solution to these limitations~\citep{Battaglia2016, Battaglia2018, Sanchez2020}. Graph neural network (GNN)-based simulators (GNSs) are one such learned physics simulator that represents the domain as a graph with vertices and edges and learns the local interactions between the vertices, which are critical in governing the dynamics of physical systems. By learning the interaction law, GNS demonstrates high predictive accuracy and generalization ability in modeling granular flows beyond the training data~\citep{choi2023_gns-column}. Since GNS is based on neural network foundations, it is inherently differentiable through automatic differentiation (AD), making it suitable for gradient-based optimization~\citep{allen2022physical, zhao2022-gns_pde_solver}.

Automatic Differentiation (AD) \citep{baydin2018ad} uniquely bridges the gap between the computational rigor required for differentiating complex functions and the practical necessity of efficiency in neural network-based systems like GNS. In contrast to traditional differentiation methods such as manual, numerical, or symbolic differentiation, AD systematically breaks down functions into simple differentiable operations. This process enables precise derivative calculations without the approximation errors inherent in numerical methods or the complexity escalation typical of symbolic differentiation. AD-based simulators are increasingly popular in solving inverse problems in fluid dynamics~\citep{wang2023inverse} and optimizing control for robotics~\citep{hu2019chainqueen}.

In the context of GNS, we leverage reverse-mode AD to compute the gradients required for solving the inverse problem. Reverse-mode AD, commonly known as backpropagation \citep{hecht1992backprop_theory, lecun1988backprop} in neural networks, calculates gradients by first performing a forward pass through the network, storing intermediate values. It then executes a backward pass, efficiently propagating gradients from the output back to the inputs. This approach is particularly suited for GNS, where the number of parameters (inputs) significantly exceeds the number of outputs (e.g., a loss function). Using the AD gradient, we optimize the input parameters through gradient-based methods. By leveraging the computational efficiency and differentiability of GNS through AD, our method offers an efficient framework for solving inverse and design problems in granular flows. 

A key challenge in applying AD to GNS, especially in simulations with extended temporal scales, is the extensive GPU memory requirement~\citep{zhao2022-gns_pde_solver}. AD necessitates tracking all intermediate computations of GNS over the entire simulation for gradient evaluation. To mitigate this issue, we employ gradient checkpointing proposed by~\cite{chen2016training}. This approach significantly reduces the memory demand by strategically storing only key intermediate computational steps during gradient evaluation. While this technique conserves memory, it requires re-evaluating certain forward computations during the backward pass for gradient evaluation. This hybrid approach allows for monitoring gradients across thousands of particle trajectories over prolonged simulation durations without overwhelming GPU memory, albeit with an increased computational burden.

\begin{figure}[]
    \centering
    \includegraphics[width=0.9\textwidth]{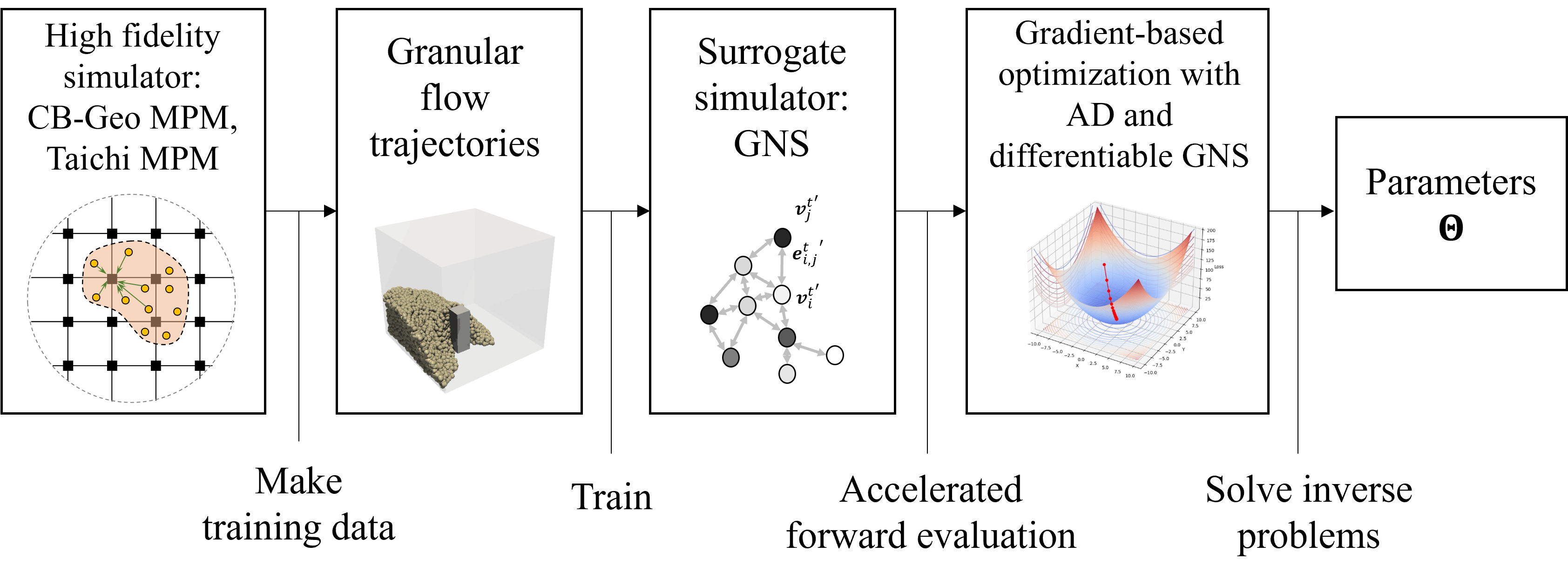}
    \caption{Overview of the research.}
    \label{fig:overview}
\end{figure}

We introduce a novel framework for inverse analysis in granular flows, leveraging the automatic differentiation capabilities of GNS with gradient-based optimization. We demonstrate our gradient-based optimization framework on inverse problems and design applications in granular flows. The overview of the research is shown in \cref{fig:overview}. We first create granular flow trajectories using the material point method (MPM)~\citep{soga2016,kumar2019scalable, hu2018mlsmpmcpic}. The trajectories are stored as NPZ files (compressed NumPy file format) and are used to train GNS. We then use GNS as a surrogate to accelerate the forward evaluation of granular flow runouts. To solve the inverse problem, we leverage the reverse-mode Automatic Differentiation of GNS (AD-GNS) nature of GNS to calculate the derivatives of the target with respect to the input parameters. Finally, we use the gradients to optimize the parameters of interest based on a target iteratively. In the following sections, we describe the details of the proposed differentiable AD-GNS and gradient-based optimization approaches for solving inverse problems.

\section{Problem statement} \label{sec:problem_statement}
\begin{figure}[]
    \centering
    \includegraphics[width=0.9\textwidth]{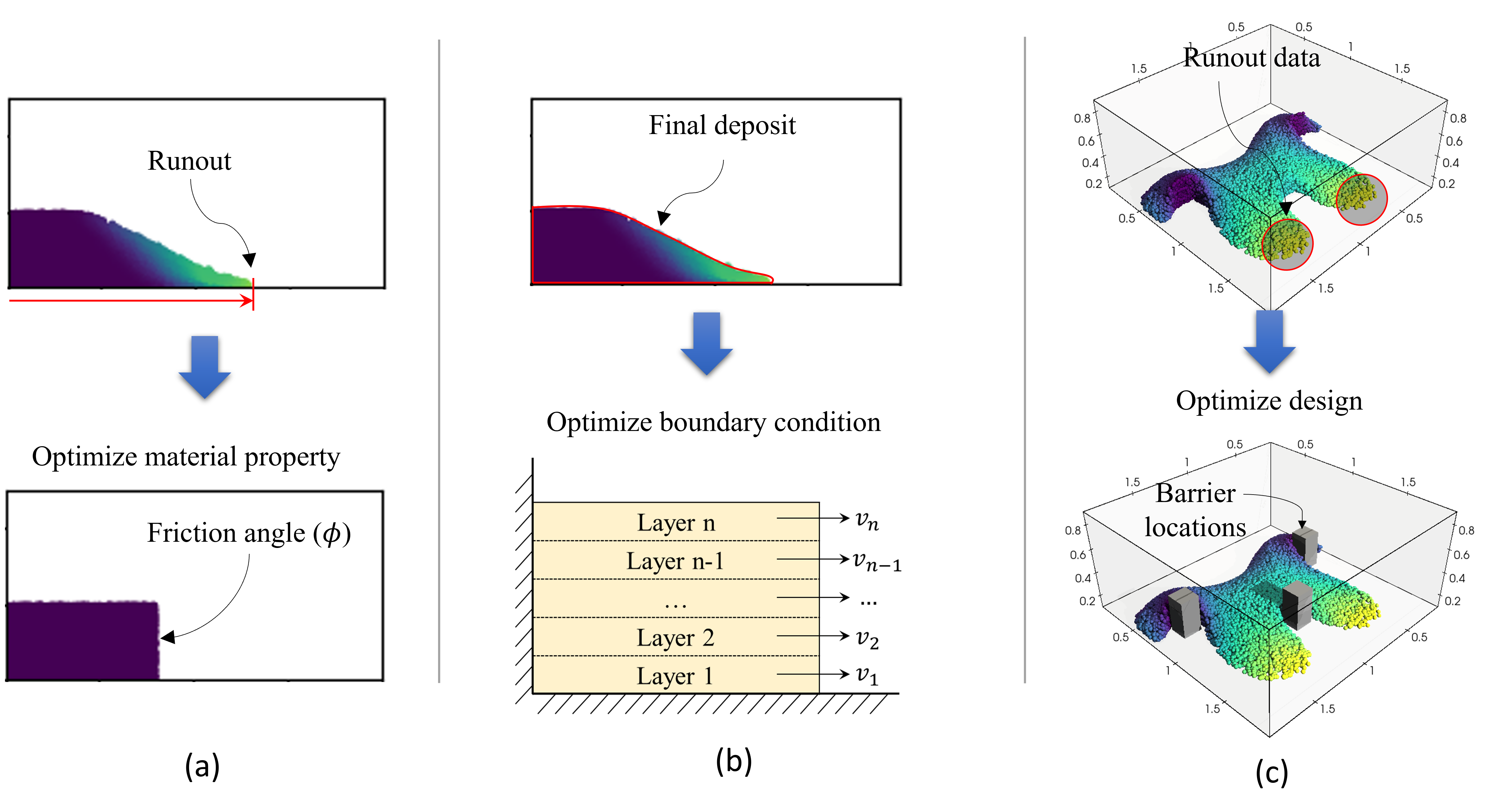}
    \caption{Illustrative representation of inverse problems in granular media: (a) single parameter inverse of determining material property based on the final runout, (b) evaluating the initial boundary conditions (initial velocity) based on the final runout profile, and (c) optimizing the location of barriers based on the target runout distribution.}
    \label{fig:problem_statement}
\end{figure}

Inverse problems involve determining the underlying parameters from observed outcomes. We investigate three inverse problems in granular flows using the differentiable GNS framework, as shown in~\cref{fig:problem_statement}. We evaluate (a) single parameter inverse of determining material property based on the final runout, (b) multi-parameter inverse of evaluating the initial boundary conditions, i.e., identify the initial velocities of the $n$-layered granular mass given the final runout profile, and (c) optimizing the location of barriers based on the target runout distribution, such as the centroid of the final deposit. In these problems, we optimize by calculating the gradient of the input parameter(s) with respect to the observed output target.

\begin{figure}
    \centering
    \includegraphics[width=0.9\textwidth]{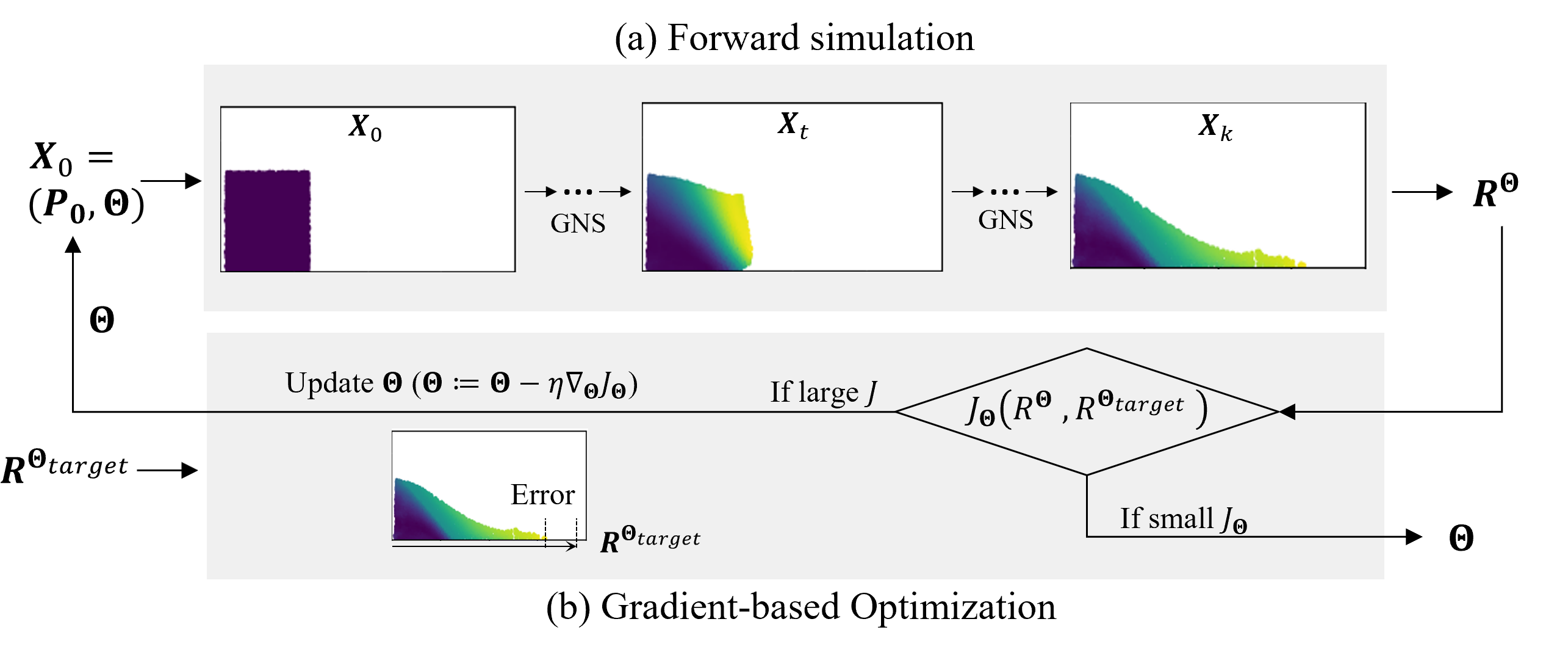}
    \caption{Schematic diagram for differentiable GNS with AD (AD-GNS) for inverse analysis. It estimates the parameters $\boldsymbol{\Theta}$ in granular column flow given a target runout distance $R^{\boldsymbol{\Theta_{target}}}$.}
    \label{fig:inverse_schematic}
\end{figure}

\Cref{fig:inverse_schematic} illustrates our framework to solve inverse problems in granular flows using differentiable GNS with AD (AD-GNS). Based on an initial set of input properties, we use the GNS to solve the forward problem of granular flow (\cref{fig:inverse_schematic}a). Given a target parameter, we evaluate a loss function ($J_{\boldsymbol{\Theta}}$), e.g., the error between the observed runout ($R^{\boldsymbol{\Theta}}$) and the target runout ($R^{\boldsymbol{\Theta_{target}}}$). Using reverse-mode AD, we compute the gradient of the loss function with respect to the optimization parameter $\nabla_{\boldsymbol{\Theta}} J_{\boldsymbol{\Theta}}$. We then iteratively update the optimization parameter ($\boldsymbol{\Theta}$) using a gradient-based optimization technique with a learning rate $\eta$ (\cref{fig:inverse_schematic}b). This process continues until the loss decreases below a threshold, thus achieving the target response. We demonstrate GNS's ability to solve inverse problems by optimizing material properties, initial boundary conditions, and design of barriers (\cref{fig:problem_statement}). In the following sections, we introduce GNS and gradient-based optimization.

\section{Methods}
\label{sec:method}

\subsection{Graph neural network-based forward simulator (GNS)}
GNS is an efficient surrogate for high-fidelity simulations of granular media. GNS discretizes the domain into a set of vertices $V$ on a graph ($G$), with each vertex representing an individual particle or a region and its properties, and the edges $E$ represent the interaction between these regions. GNS takes the current state of the domain $\boldsymbol{X}_t$ at time $t$ and returns its next state $\boldsymbol{X}_{t+1}$ (i.e., $GNS: \ \boldsymbol{X}_{t} \rightarrow \boldsymbol{X}_{t+1}$). $\boldsymbol{X}_t$ contains information about particles' position, velocity, distance to boundaries, and material properties. A surrogate simulation of granular flow involves running GNS through $k$ timesteps predicting from the initial state $\boldsymbol{X}_0$ to $\boldsymbol{X}_k$ (i.e., $\boldsymbol{X}_0 \rightarrow \boldsymbol{X}_1 \rightarrow \ldots \rightarrow \boldsymbol{X}_k$) sequentially. We call this successive forward GNS prediction the ``rollout''. In the following paragraphs, we briefly introduce the structure of GNS. For more details about GNS, please see~\cite{Sanchez2020} and \cite{choi2023_gns-column}. 

\begin{figure}[]
    \centering
    \includegraphics[width=0.7\textwidth]{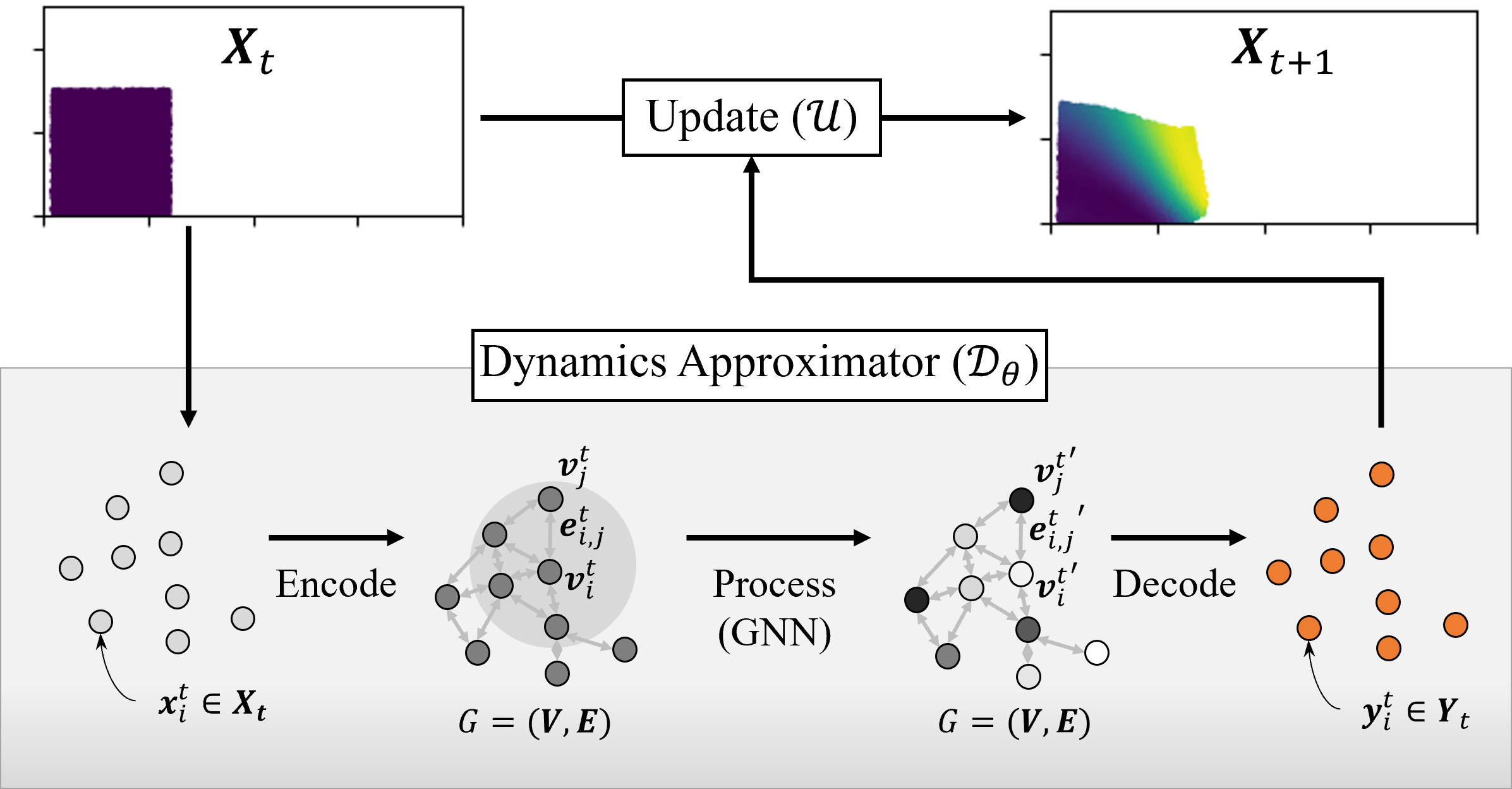}
        \caption{Components of graph neural network (GNN)-based learned simulator (GNS)}
\label{fig:GNS}
\end{figure}

GNS consists of two components (\cref{fig:GNS}): dynamics approximator $\mathcal{D}_{\theta}$, which is a learned function approximator parameterized by $\theta$, and update function  $\mathcal{U}$.  $\mathcal{D}_{\theta}$ has an encoder-processor-decoder architecture. The encoder converts $\boldsymbol{X}_t$ to a latent graph $G=(\boldsymbol{V}, \boldsymbol{E})$ representing the state of particle interactions. The processor propagates information between vertices by passing messages along edges and returns an updated graph $G'=(\boldsymbol{V}', \boldsymbol{E}')$. In the physics simulations, this operation corresponds to energy or momentum exchange between particles. The decoder extracts dynamics $\boldsymbol{Y}_t$ of particles from the updated graph. The update function uses the dynamics to update the current state to the next state of material points ($\boldsymbol{X}_{t+1}=\mathcal{U}(\boldsymbol{X}_t, \boldsymbol{Y}_t)$). The update function $\mathcal{U}$ is analogous to the explicit Euler integration in numerical solvers. Hence, in our GNS, $\boldsymbol{Y}_t$ corresponds to the second derivative of the current state, i.e., acceleration. 

The encoder and decoder operations employ multi-layered perceptrons (MLPs), while the processor comprises GNN, which contains the learnable parameters $\theta$. We train this $\theta$ to minimize the mean squared error between the ground truth accelerations of the particles $\boldsymbol{A}_t$ and the predicted dynamics $\boldsymbol{Y}_t$ given the current state of particles $\boldsymbol{X}_t$ as the input. In our study, we use the material point method (MPM)~\citep{soga2016,kumar2019scalable} to generate the training dataset of particles (material point) positions $\boldsymbol{X}_t$ and acceleration $\boldsymbol{A}_t$. 

We leverage PyTorch's Distributed Data Parallelism (DDP) framework to enable multiple GPU parallelization. This approach demonstrates almost linear scaling performance \citep{kumar2022gns, designsafe-choi-gns}. By employing DDP, we can distribute the workload associated with large graph structures across multiple GPUs with batching, thus facilitating the operation of GNS with a distributed computational burden without compromising the speed.

\subsubsection{Training data}
We prepare two types of datasets: two-dimensional granular flows (Flow2D) and three-dimensional granular flows interacting with obstacles (Obstacle3D). \Cref{table:train_data} shows the details of the simulation configurations.

We train the GNS model for the two-dimensional domain (\cref{fig:problem_statement}a and \cref{fig:problem_statement}b) on the Flow2D dataset. It includes 385 square-shaped granular mass flow trajectories in a two-dimensional box boundary. The appendix section provides a visual example in \cref{fig:train_data_2d}. We use five different friction angles ($\phi=15 \degree, 22.5 \degree, 30 \degree, 37.5 \degree, 45 \degree$), where each friction angle has 77 simulation datasets. Each simulation has different initial configurations regarding the size of the square granular mass, position, and velocity. We use the explicit time integration method using CB-Geo MPM code \citep{kumar2019scalable} to generate the training data. 

We train the 3D GNS (see \cref{fig:problem_statement}c) on the Obstacle3D dataset. It includes 1000 trajectories of granular mass interacting with barriers in a three-dimensional box boundary. The appendix section provides a visual example in \cref{fig:train_data_3d}. The initial geometry of the cuboid-shaped granular mass varies from 0.25 to 0.80 $m$ and is excited with different initial velocities. For each simulation, the granular mass interacts with one to three cuboid-shaped barriers whose width and length vary from 0.10 to 0.13 $m$ with a height of 0.3 $m$ in the $1.0\times1.0\times1.0 \ m$ domain. We use the explicit time integration method using Moving Least Squares (MLS)-MPM code (Taichi MPM) \citep{hu2018mlsmpmcpic} to generate the training data. See \ref{sec:appendix-performance} for the forward simulation performance of GNS.

We use two different MPM software, CB-Geo MPM (2D simulations) and Taichi MPM (3D simulations), to get the best of both worlds. While CB-Geo MPM can reasonably simulate 2D granular flow, a 3D MPM code with rigid body dynamics was needed to model the barrier problems – hence, we use Taichi MPM code.

\begin{table}[]
\caption{Details of the Material Point Method (MPM) simulation geometries and properties used for generating the training datasets (reproduced after \citep{choi2023_gns-column})}
\label{table:train_data}
\resizebox{\textwidth}{!}{%
\begin{tabular}{@{}llll@{}}
\toprule
\multicolumn{2}{c}{\multirow{2}{*}{Property}} & Datasets &  \\ \cmidrule(l){3-4} 
\multicolumn{2}{c}{} & Flow2D & Obstacle3D \\ \midrule
\multicolumn{2}{l}{Simulation boundary} & 1.0×1.0 $m$ & 1.0×1.0×1.0 $m$ \\
\multicolumn{2}{l}{MPM element length} & 0.01×0.01 $m$ & 0.03125×0.03125×0.03125 $m$ \\
\multicolumn{2}{l}{Material point configuration} & 40,000 $points/m^2$ & 262,144 $points/m^3$ \\
\multicolumn{2}{l}{Granular mass geometry} & 0.2×0.2 to 0.4×0.4 $m$ & 0.25 to 0.80  $m$ for each dimension \\
\multicolumn{2}{l}{Max. number of particles} & 6.4K & 17K \\
\multicolumn{2}{l}{Barrier geometry} & None & 0.10×0.3×0.10 to 0.13×0.3×0.13 $m$ \\
\multicolumn{2}{l}{Simulation duration (\# of timesteps)} & 1.0 s (400) & 0.875 s (350) \\ \midrule
\multirow{7}{*}{Material property} & Model & Mohr-Coulomb & Mohr-Coulomb \\
 & Density & 1,800 $kg/m^3$ & 1,800 $kg/m^3$ \\
 & Young's modulus & 2 $MPa$ & 2 $MPa$ \\
 & Poisson ratio & 0.3 & 0.3 \\
 & Friction angle & 15, 22.5, 30, 37.5, 45 $\degree$ & 35$\degree$ \\
 & Cohesion & 0.1 $kPa$ & None \\
 & Tension cutoff & 0.05 $kPa$ & None \\ \bottomrule
\end{tabular}%
}
\end{table}

\subsection{Differentiable GNS}\label{sec:Differentiable GNS}
GNS built using neural networks is fully differentiable \citep{allen2022physical} using reverse-mode AD. Reverse-mode AD computes the gradient of the output of GNS with respect to its inputs by constructing a computational graph of the simulation and propagating a gradient through the graph. Using the gradient, we can solve inverse problems using gradient-based optimization (\Cref{fig:inverse_schematic}b). This gradient-based optimization allows very efficient parameter updates, particularly in high-dimensional space, as it uses the gradient information to move the optimization parameters $\boldsymbol{\Theta}$ towards the optimum solution \citep{allen2022physical, dhara2023fwi}. Here, we describe the details of reverse-mode AD, followed by the gradient-based optimization in the later section (\cref{sec:grad-based-opt}).

Reverse-mode AD \citep{baydin2018automatic} offers a highly efficient method for calculating the accurate gradients of functions. Unlike numerical differentiation, which is approximate and computationally expensive, or symbolic differentiation, which can be unwieldy for complex functions, reverse-mode AD efficiently computes analytical gradients. 

\begin{figure}[]
    \centering
    \includegraphics[width=0.8\textwidth]{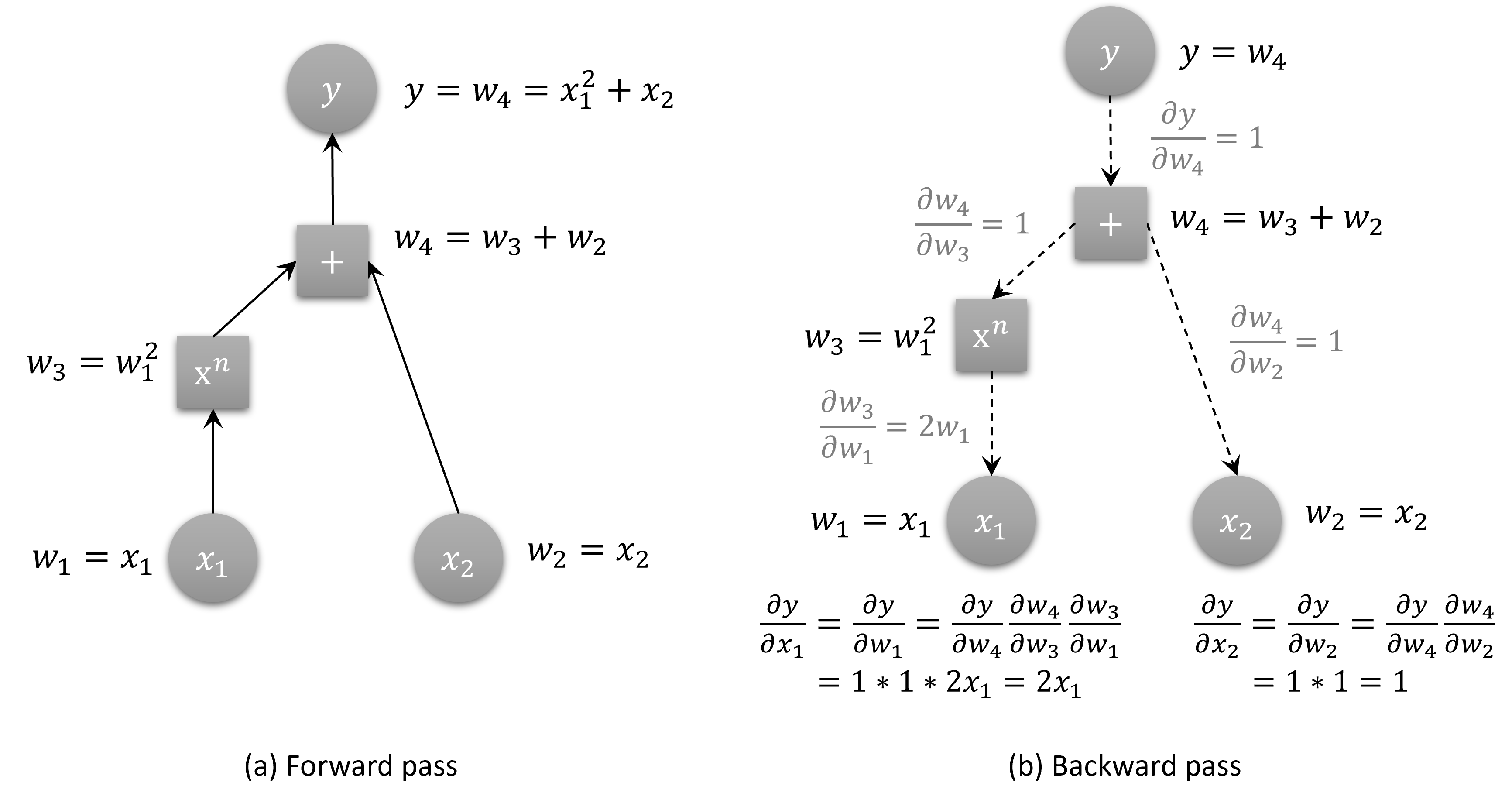}
    \caption{Schematics of a Directed Acyclic Graph (DAG) of reverse-mode automatic differentiation (AD) of an example function.}
    \label{fig:reverse_mode_ad}
\end{figure}

Reverse-mode AD leverages computational graphs (\cref{fig:reverse_mode_ad}) to represent complex functions and efficiently and accurately compute their derivatives. A computational graph is directed where each node represents an elementary operation or variable, and edges indicate the computation flow. The combination of the nodes and edges makes computational graphs represent complex functions. 

To compute the gradient, reverse-mode AD conducts two steps: forward and backward pass. In the forward pass (\cref{fig:reverse_mode_ad}a), reverse-mode AD evaluates the function's output given the inputs. Starting from the input nodes, which represent the input variables ($x_1$ and $x_2$), the graph is traversed to compute the output of the function. This step lays the groundwork for the differentiation process by establishing the relationships between all the operations involved in computing the function. 

The next step is backward pass (\cref{fig:reverse_mode_ad}b), known as reverse mode or backpropagation, which is the core process of reverse-mode AD to compute the gradient with respect to the inputs of the function ($x_1$ and $x_2$). This phase starts from the output of the computational graph ($y$) and works backward. As the process moves backward through the graph, reverse-mode AD computes what is known as a local derivative—the derivative of that node's operation with respect to its immediate input. In \cref{fig:reverse_mode_ad}b, this corresponds to $\partial{y}/\partial{w_4}$, $\partial{w_4}/\partial{w_3}$, $\partial{w_3}/\partial{w_1}$, and $\partial{w_4}/\partial{w_2}$, denoted beside the gray dashed edges. By representing the simulation as a computational graph, we can pre-compute the derivatives, thus improving computational efficiency in gradient evaluation. We employ the chain rule, which combines every local derivative at each node from the output to the input node, returning the gradient of output with respect to the graph's input ($\partial{y}/\partial{x_1}$, $\partial{y}/\partial{x_2}$). In \cref{fig:reverse_mode_ad}b, this process is described below the nodes $x_1$ and $x_2$.

The forward pass of the GNS is described as:

\begin{equation}\label{eq:GNS-forward}
\boldsymbol{X}_{t+1} = GNS(\boldsymbol{X}_t) \quad \mathrm{for} \ t \in \{ 0, 1, \ldots, k-1 \}
\end{equation}

\noindent where $\boldsymbol{X}_t$ is the current state of the system at time $t$, and GNS is the learned function approximator that predicts the next state $\boldsymbol{X}_{t+1}$ given the current state $\boldsymbol{X}_t$. The forward pass involves predicting the trajectory for $k$ timesteps to simulate the evolution of the system from the initial state $\boldsymbol{X}_0$ to $\boldsymbol{X}_k$.

For the inverse analysis, we define an objective function (loss) $J_{\boldsymbol{\Theta}}$ that needs to be minimized with respect to the parameters ${\boldsymbol{\Theta}}$. The objective function depends on the output of GNS after rolling out for $k$ timesteps, starting from the initial state $\boldsymbol{X}_0$.

\begin{equation}
    J_{\boldsymbol{\Theta}} = J(\boldsymbol{X}_k)
\end{equation}

\noindent where $\boldsymbol{X}_k$ is obtained by recursively applying $GNS()$ for $k$ timesteps:

\begin{align*}
    \boldsymbol{X}_1 &= GNS(\boldsymbol{X}_0; {\boldsymbol{\Theta}})\\
\boldsymbol{X}_2 &= GNS(\boldsymbol{X}_1; {\boldsymbol{\Theta}})\\
\vdots \\
\boldsymbol{X}_k &= GNS(\boldsymbol{X}_{k-1}; {\boldsymbol{\Theta}})
\end{align*}

\noindent We can compactly represent this recursive forward pass as:
\begin{equation}
\boldsymbol{X}_k = GNS^k(\boldsymbol{X}_0; {\boldsymbol{\Theta}})
\end{equation}

\noindent where $GNS^k$ denotes applying $GNS()$ recursively for $k$ timesteps.

To compute the gradient of $J_{\boldsymbol{\Theta}}$ with respect to ${\boldsymbol{\Theta}}$ using reverse mode AD, we construct the computational graph for the entire forward pass $GNS^k$ and then apply the chain rule:
\begin{equation}
    \nabla_{\boldsymbol{\Theta}} J_{\boldsymbol{\Theta}} = \frac{\partial J}{\partial \boldsymbol{X}_k} \frac{\partial \boldsymbol{X}_k}{\partial \boldsymbol{\Theta}}
\end{equation}

\noindent where $\frac{\partial \boldsymbol{X}_k}{\partial \boldsymbol{\Theta}}$ is computed by recursively applying the chain rule through the entire trajectory:

\begin{equation}\label{eq:GNS-backward}
\frac{\partial \boldsymbol{X}_{k}}{\partial \boldsymbol{\Theta}} = \frac{\partial \boldsymbol{X}_{k}}{\partial \boldsymbol{X}_{k-1}} \ldots \frac{\partial \boldsymbol{X}_{1}}{\partial \boldsymbol{X}_{0}} \frac{\partial \boldsymbol{X}_{0}}{\partial \boldsymbol{\Theta}}
\end{equation}

This reverse mode AD computes the exact gradients $\nabla_{\boldsymbol{\Theta}} J_{\boldsymbol{\Theta}}$ by propagating the gradients from the final state $\boldsymbol{X}_k$ back to the initial state $\boldsymbol{X}_0$ and then to the parameters $\boldsymbol{\Theta}$, while leveraging the intermediate states $\boldsymbol{X}_1, \boldsymbol{X}_2, \cdots, \boldsymbol{X}_{k-1}$ computed during the forward pass.

Note that storing all these intermediate states for long trajectories (large $k$) can be memory-intensive, which is where gradient checkpointing (\cref{sec:grad_check}) is employed to reduce the memory footprint by selectively storing and recomputing intermediate states.

Unlike the finite difference method, which needs multiple function evaluations to estimate the derivative of each parameter, reverse-mode AD computes precise analytic gradients in a single forward pass. This makes reverse-mode AD particularly advantageous in gradient-based optimization, especially in large parameter space scenarios. The parameters can be efficiently updated towards the optimum based on the gradient information obtained from reverse-mode AD. In the following section, we elaborate on the gradient-based optimization with reverse-mode AD. 

\subsection{Gradient-based optimization}\label{sec:grad-based-opt}
\Cref{eq:general_gd} shows the general concept of a simple gradient descent optimization scheme. Gradient descent minimizes the discrepancy between the simulator output and the target data, defined as an objective function $J_{\boldsymbol{\Theta}}$, by iteratively updating the model's parameter set $\boldsymbol{\Theta}$. The gradient of the objective function provides the information about the most effective update direction with respect to the parameters, $\nabla_{\boldsymbol{\Theta}} J_{\boldsymbol{\Theta}}$. Here, $\eta$ is the learning rate. 

\begin{equation} \label{eq:general_gd}
    \boldsymbol{\Theta} := \boldsymbol{\Theta} - \eta \cdot \nabla_{\boldsymbol{\Theta}} J_{\boldsymbol{\Theta}}
\end{equation}

The learning rate or step size in an optimization algorithm determines the increment at each iteration while traversing toward a minimum of a loss function; this learning rate or step size $\eta$, multiplied by the gradient $\nabla_{\boldsymbol{\Theta}} J_{\boldsymbol{\Theta}}$, controls the aggressiveness of parameter updates in a gradient descent optimization of multi-variable function $J_{\boldsymbol{\Theta}}$ (see \cref{eq:general_gd}). A high learning rate can lead to instability in the optimization process, with large steps causing overshooting of the minimum or even divergence. Conversely, a small learning rate will result in slow convergence, as the model takes minuscule steps and requires excessive iterations to reach a solution. Finding the optimal learning rate is largely experimental. We start with a small learning rate (e.g., 0.001), observe the parameter updates in the initial iterations, and manually vary the learning rate to achieve convergence in a reasonable time. We can also employ adaptive learning rates or learning rate schedulers that decrease the rate over time for stability.  Alternatively, adaptive optimizers like adaptive movement estimation algorithm (ADAM) \citep{kingma2014adam} or root mean squared propagation (RMSprop) automate some of the learning rate adjustment process.

The differentiable simulator computes the gradient $\nabla_{\boldsymbol{\Theta}} J_{\boldsymbol{\Theta}}$ using reverse-mode AD to update the parameter set $\boldsymbol{\Theta}$ at each optimization step. Since GNS is a differentiable simulator, we can seamlessly utilize the AD tools provided by frameworks such as PyTorch to calculate the gradient efficiently. 

This approach offers several advantages. First, it provides a systematic way of handling complex, nonlinear inverse problems where analytical solutions are not feasible. Second, the efficiency of reverse-mode AD in differentiable simulators enables handling high-dimensional parameter spaces and complex relationships between parameters and outputs. Finally, gradient-based optimization often leads to faster convergence to a solution compared to other sampling-based optimization methods, especially in large-scale problems \citep{allen2022physical}. 

We use the adaptive movement estimation algorithm (ADAM)~\citep{kingma2014adam} for high-dimensional parameter optimization. ADAM incorporates a dynamic adaptive learning rate for each parameter using the moments of the gradients. ADAM accelerates the parameter updates in the right direction, thus avoiding local minima and adapting the learning rate for each parameter, making it less sensitive to the scale of the gradients. 

Specifically, the parameters are updated as:

\begin{equation}\label{eq:adam-eq}
\boldsymbol{\Theta}_{i} = \boldsymbol{\Theta}_{i-1} - \frac{\eta}{\sqrt{\hat{v}_i} + \epsilon} \hat{m}_i
\end{equation}

\noindent where $\boldsymbol{\Theta}$ is the parameter set with the optimization step $i$, and $\epsilon$ is a small scalar for numerical stability. The first and second moments ($m_i$ and $v_i$) of the gradients are updated as:

\begin{align}\label{eq:adam-moment}
m_i &= \beta_1 m_{i-1} + (1 - \beta_1) g_i \\
v_i &= \beta_2 v_{i-1} + (1 - \beta_2) g_i^2
\end{align}

\noindent where $g_i$ represents the gradient at step $i$, $\beta_1$ and $\beta_2$ are decay rates that control the exponential decay of these moving averages. Early in the training process, the variables $m_i$ and $v_i$ are initially biased toward zero because they start from zero. To address this bias, a correction is applied as:

\begin{align}\label{eq:adam-bias}
\hat{m}_i &= \frac{m_i}{1 - \beta_1^i} \\
\hat{v}_i &= \frac{v_i}{1 - \beta_2^i}
\end{align}

This adaptive learning rate is particularly effective optimization in high-dimensional parameter spaces, which entails sparse and varied gradient scales. ADAM adapts the learning rate for each parameter based on its historical gradients, which helps in scenarios where different parameters have different scales or degrees of sparsity. This adaptive learning makes ADAM converge quicker and improves performance in solving multi-parameter inverse problems. We refer readers to \cite{kingma2014adam} for more technical details.

\subsubsection{Gradient checkpointing} \label{sec:grad_check}
Reverse-mode AD requires significant memory for large-scale neural networks \citep{zhao2022-gns_pde_solver}. AD necessitates storing intermediate variables in the computational graph during the forward pass. These intermediate variables should be retained during the backward pass to apply the chain rule in computing gradients. The computational graph grows for large-scale neural networks by adding more layers and parameters. Therefore, the backpropagation requires substantial memory to retain all the intermediate variables in the increasingly large computational graph.

Since GNS contains multiple MLP and GNN layers containing millions of parameters, and the entire simulation even entails the accumulation of GNS ($\boldsymbol{X}_t\rightarrow \boldsymbol{X}_{t+1}$) for $k$ steps, computing  $\nabla_{\boldsymbol{\Theta}} J_{\boldsymbol{\Theta}}$ using reverse-mode AD requires extensive memory capacity. We found that conducting reverse-mode AD for entire simulation timesteps, which includes more than hundreds of steps, is not feasible in the currently available GPU memory capacity (40 GB). The forward pass fails in 3 to 4 timesteps due to the lack of memory when simulating granular flows with about 3K particles. To overcome this limitation, we employ gradient checkpointing.

Gradient checkpointing \citep{chen2016training} is a technique to mitigate the prohibitive memory demands of storing all intermediate states within a computational graph during reverse-mode AD. In the specific setting of optimizing parameters with GNS, we face the challenge of an extensive computational graph that spans both GNN message-passing layers and simulation timesteps. Checkpointing addresses this by strategically saving a subset of the intermediate states (checkpoints) and discarding the rest. During the backward pass, any required states that are not checkpoints are recomputed on the fly from the most recent checkpoint. This selective storage significantly reduces the peak memory requirement at the expense of some recomputation. 

Let's consider the forward pass of GNS over $k$ timesteps: 
\begin{equation}
    \boldsymbol{X}_0 \rightarrow\boldsymbol{X}_1\rightarrow\boldsymbol{X}_2 \cdots\rightarrow\boldsymbol{X}_k
\end{equation}
During the forward pass, instead of storing all the intermediate states $\boldsymbol{X}_0, \boldsymbol{X}_1, \boldsymbol{X}_2, \cdots,\boldsymbol{X}_{k-1}$, gradient checkpointing selectively stores only a subset of these states, called checkpoints, at specific timesteps $C \in \{ 0, c_1, c_2, \cdots, c_m, k \}$ as:

\begin{equation}
    \boldsymbol{X}_{i} = \text{GNS}(\boldsymbol{X}_{i-1}); \quad
    \left\{
    \begin{array}{ll}
        \text{store} & \leftarrow \boldsymbol{X}_{i} \quad \forall i \in C, \\
        \text{discard} & \leftarrow \boldsymbol{X}_{i} \quad \forall i \notin C
    \end{array}
    \right.
\end{equation}

\noindent During the backward pass, those intermediate states between $\boldsymbol{X}_{i-1}$ and $\boldsymbol{X}_i$ $\forall i \notin C$ are recomputed to evaluate the local derivatives. The reverse mode AD with gradient checkpoints is computed as follows:

\begin{enumerate}
    \item \textbf{Initial gradient computation at final state:} Start by computing the gradient of the loss function $J$ with respect to the final state $\boldsymbol{X}_k$: $\frac{\partial J}{\partial \boldsymbol{X}_k}$
    This gradient is derived from the final output of the GNS model, based on how the loss $J$ is affected by the last computed state $\boldsymbol{X}_k$.

    \item \textbf{Backpropagate through GNS from $ \boldsymbol{X}_k $ to last checkpoint $ \boldsymbol{X}_{c_m} $:} \\
    Use the gradient from the final state to backpropagate through the operations performed from the last checkpoint to the final state:
    \begin{equation*}
    \frac{\partial J}{\partial \boldsymbol{X}_{c_m}} = \left(\frac{\partial J}{\partial \boldsymbol{X}_k}\right) \cdot \left(\frac{\partial \boldsymbol{X}_k}{\partial \boldsymbol{X}_{c_m}}\right)
    \end{equation*}
    Here, $ \frac{\partial \boldsymbol{X}_k}{\partial \boldsymbol{X}_{c_m}} $ represents the gradient of the output at step $ k $ with respect to the state at the last checkpoint $ c_m $, encompassing all intermediate transformations.

    \item \textbf{Backpropagate successively through each checkpoint:} \\
    For each checkpoint $ c_m $, starting from the last stored checkpoint $ c_m $ and moving backward through earlier checkpoints $ c_{m-1}, c_{m-2}, \ldots $, compute:
    \begin{equation*}
    \frac{\partial J}{\partial \boldsymbol{X}_{c_{(m-1)}}} =  
    \left(\frac{\partial J}{\partial \boldsymbol{X}_{c_m}}\right) \cdot 
    \left(\frac{\partial \boldsymbol{X}_{c_m}}{\partial \boldsymbol{X}_{c_{(m-1)}}}\right)
    \end{equation*}
    
     Each term $\frac{\partial \boldsymbol{X}_{c_m}}{\partial \boldsymbol{X}_{c_{(m-1)}}}$ captures the gradient of the checkpointed state with respect to its preceding checkpoint, reflecting the operations of $\text{GNS}$ executed between these points.

     \item \textbf{Recompute non-stored states as needed:} \\
     Assuming a non-checkpointed intermediate state $i$ between checkpoints $c_{(m-1)}$ and $c_{(m-2)}$, where $c_{(m-2)} < i < c_{(m-1)}$ and $\boldsymbol{X}_i$ is not stored during the forward pass. We handle the backward pass as:

\begin{enumerate}
   \item \textbf{Recompute intermediate states:} 
  If $\boldsymbol{X}_i$ is an intermediate state that was not stored, recomputation from the last stored state $\boldsymbol{X}_{c_{(m-2)}}$ is necessary:
  \begin{equation*}
  \boldsymbol{X}_i = \text{GNS}^{i \leftarrow c_{(m-2)}}(\boldsymbol{X}_{c_{(m-2)}})
  \end{equation*}
  This step involves executing the forward function $\text{GNS}$ iteratively from $c_{(m-2)}$ to $i$, reconstructing each $\boldsymbol{X}_i$ sequentially up to $\boldsymbol{X}_{c_{(m-1)}}$.

  \item \textbf{Backpropagate gradient through $\boldsymbol{X}_i$:}
  With $\boldsymbol{X}_i$ recomputed, the gradient from $\boldsymbol{X}_{c_{(m-1)}}$ is propagated back:
  \begin{equation*}
  \frac{\partial J}{\partial \boldsymbol{X}_i} = \left(\frac{\partial J}{\partial \boldsymbol{X}_{c_{(m-1)}}}\right) \cdot \left(\frac{\partial \boldsymbol{X}_{c_{(m-1)}}}{\partial \boldsymbol{X}_i}\right)
  \end{equation*}
  We compute the gradients for each $\boldsymbol{X}_j$ (for $j = k$ to $c_{(m-1)}$) during this backpropagation process.

  \item \textbf{Gradient at checkpoint $c_{(m-2)}$:}
  We continue the gradient propagation back to $\boldsymbol{X}_{c_{(m-2)}}$:
  \begin{equation*}
  \frac{\partial J}{\partial \boldsymbol{X}_{c_{(m-2)}}} = \left(\frac{\partial J}{\partial \boldsymbol{X}_i}\right) \cdot \left(\frac{\partial \boldsymbol{X}_i}{\partial \boldsymbol{X}_{c_{(m-2)}}}\right)
  \end{equation*}
  This step involves recalculating each partial derivative $\partial \boldsymbol{X}_j/\partial \boldsymbol{X}_{j-1}$ during the recomputation phase.
\end{enumerate}

    \item \textbf{Finalize gradients for initial state and parameters:} \\
    Once backpropagation reaches the first checkpoint $ \boldsymbol{X}_{c_1} $, complete the gradient computation for the initial state and model parameters:
    \begin{equation*}
    \frac{\partial J}{\partial \boldsymbol{X}_0} = \left(\frac{\partial J}{\partial \boldsymbol{X}_{c_1}}\right) \cdot \left(\frac{\partial \boldsymbol{X}_{c_1}}{\partial \boldsymbol{X}_0}\right)
    \end{equation*}
    \begin{equation*}
    \frac{\partial J}{\partial \boldsymbol{\Theta}} = \left(\frac{\partial J}{\partial \boldsymbol{X}_{c_1}}\right) \cdot \left(\frac{\partial \boldsymbol{X}_{c_1}}{\partial \boldsymbol{\Theta}}\right)
    \end{equation*}
     These calculations utilize the gradients at the first checkpoint to determine how initial conditions and parameters should be adjusted to minimize the loss $J$.
\end{enumerate}

By selectively storing and recomputing intermediate states, gradient checkpointing reduces the peak memory requirement during backpropagation at the cost of some additional computations during the backward pass. Choosing checkpoint locations $C$ is crucial for balancing memory savings and computational overhead. A common strategy is to place checkpoints at regular intervals (e.g., every few timesteps) or at specific points in the simulation where the computational graph is expected to be particularly large or complex. Using gradient checkpointing, we can differentiate through the entire simulation.

\section{Results of inverse analysis in granular flows}

We evaluate the performance of AD-GNS in solving the three inverse problems described in \cref{fig:problem_statement}. Note that the GNS was not trained on any of these problems to showcase the adaptability and robustness of our approach. 

\subsection{Single parameter inverse}\label{sec:friction_inverse}

Estimating the material properties of granular media resulting in a landslide or debris flow is important for planning mitigation measures. In this problem, we aim to evaluate the friction angle $\phi$ of the granular column that produces a target runout distance $d_{\phi_{target}}$  (as outlined in \cref{fig:problem_statement}a), where $d_{\phi}$ is the distance between the flow toe and the leftmost boundary. We use AD-GNS to generate a forward simulation to estimate $d_{\phi}$ for a given initial $\phi$. We evaluate a loss $J_{\phi}$ as the squared error between $d_{\phi_{target}}$ and  $d_{\phi}$ (\cref{eq:single param error}), then use AD-GNS to compute the gradient $\nabla_{\phi} J_{\phi}$. Using GD, we update the $\phi$ to find the target $\phi$ for a desired runout $d_{\phi_{target}}$ until the squared error becomes lower than 0.0005, the optimization threshold.

\begin{equation}\label{eq:single param error}
J_{\phi} = (d_{\phi} - d_{\phi_{target}})^2
\end{equation}

To test our method's versatility, we selected four granular column collapse scenarios with varying aspect ratios and friction angles not covered in our training data. These scenarios ranged from columns with a small aspect ratio (short column, $a=0.5$) to those with a large aspect ratio (tall column, $a=2.0$). Both cases have 3,200 particles in the simulation.

\begin{figure}[]
    \centering
    \includegraphics[width=1.0\textwidth]{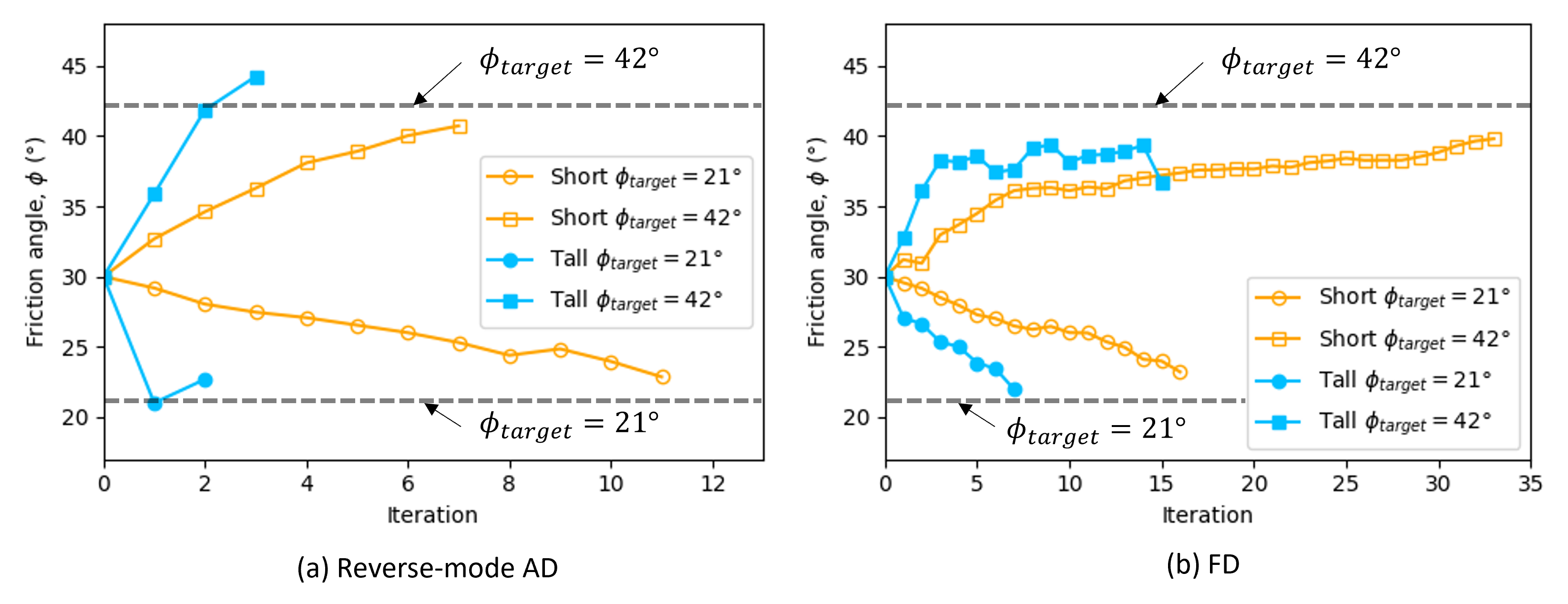}
    \caption{Optimization history of friction angle with iterations from (a) reverse-mode AD and (b) finite differentiation (FD).}
    \label{fig:optimization_phi_history}
\end{figure}

\begin{figure}[!htbp]
     \centering
     \begin{subfigure}[b]{1.0\textwidth}
         \centering
         \includegraphics[width=\textwidth]{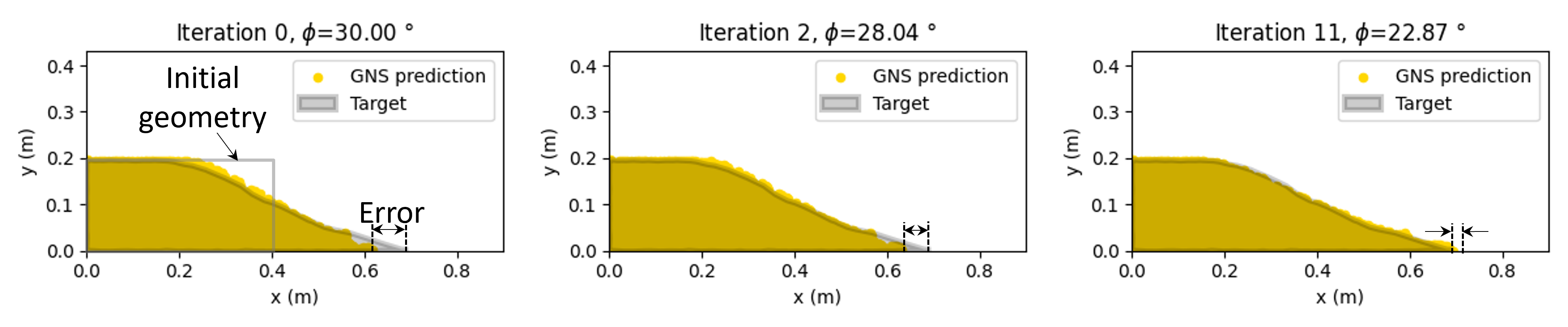}
         \caption{}
         \label{fig:vis_opt_short21}
     \end{subfigure}
     \vfill
     \begin{subfigure}[b]{1.0\textwidth}
         \centering
         \includegraphics[width=\textwidth]{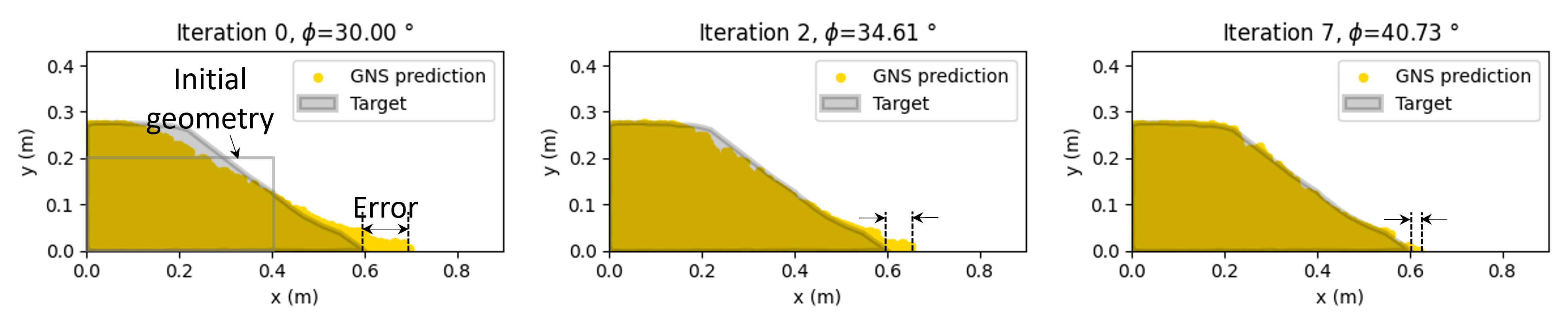}
         \caption{}
         \label{fig:vis_opt_short42}
     \end{subfigure}
     \vfill
     \begin{subfigure}[b]{1.0\textwidth}
         \centering
         \includegraphics[width=\textwidth]{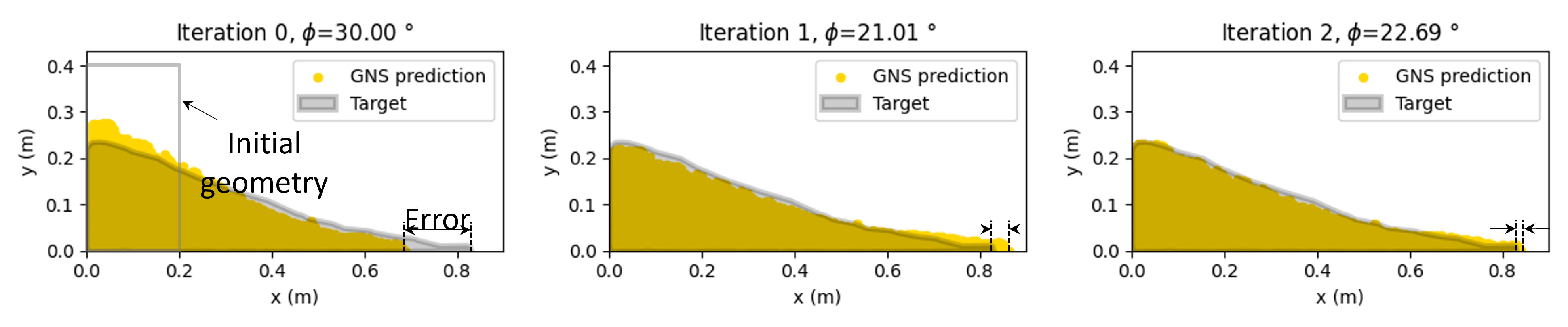}
         \caption{}
         \label{fig:vis_opt_tall21}
     \end{subfigure}
     \vfill
     \begin{subfigure}[b]{1.0\textwidth}
         \centering
         \includegraphics[width=\textwidth]{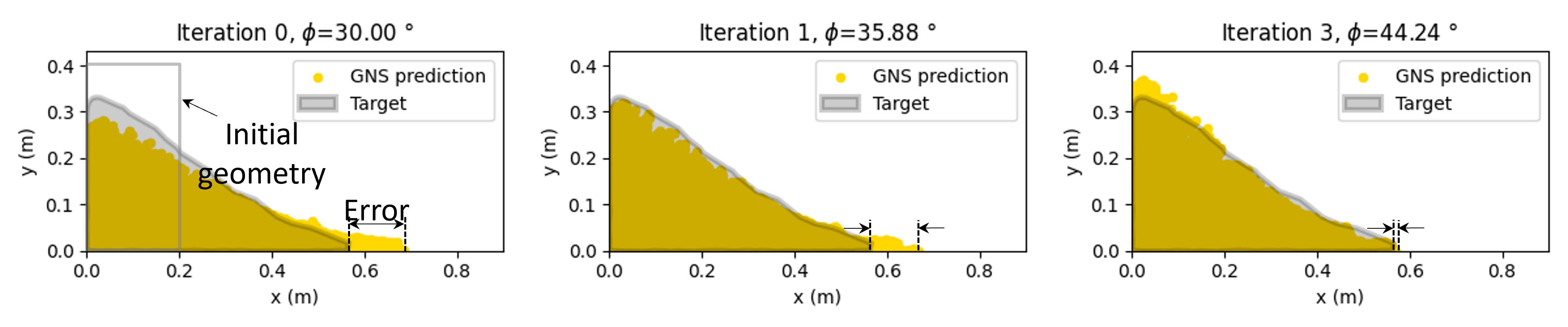}
         \caption{}
         \label{fig:vis_opt_tall42}
     \end{subfigure}
\caption{Visualization of GNS prediction during optimization by varying $\phi$: (a) short column ($a=0.5$) with $\phi_{target}=21 \degree$, (b) short column ($a=0.5$) with $\phi_{target}=42 \degree$, (c) tall column ($a=2.0$) with $\phi_{target}=21 \degree$, (d) tall column ($a=2.0$) with $\phi_{target}=42 \degree$. The yellow dots represent the final deposit predicted by GNS, and the gray shade represents the final deposit from MPM with to $\phi_{target}$}
\label{fig:visual_phi_opt}
 \end{figure}

\Cref{table:result_phi} summarizes the scenarios and results for the inversely estimated friction angles. \Cref{fig:optimization_phi_history}a shows the optimization history based on reverse-mode AD, and \cref{fig:visual_phi_opt} shows the visual progress of the optimization for all the test scenarios. We set the learning rate $\eta$ of 500 in \cref{eq:general_gd}, and the initial guess of the friction angle starts from $30 \degree$. As can be seen in \cref{fig:optimization_phi_history}a, the optimization successfully converges to the friction angles $\phi$ that are close enough to the target values $\phi_{target}$ showing errors less than 8.94\% (see \cref{table:result_phi}) in a few iterations. We also compare the predicted runout distance $d_{\phi}^{GNS}$ with the inferred $\phi$ and target runout distance $d_{\phi_{target}}$. The predictions show accurate values with a maximum error of 3.49\% (see \cref{table:result_phi}). Although the inverse estimation of $\phi$ includes a small error (about $1 \degree$ to $3 \degree$) compared to $\phi^{target}$, the runout predictions are accurate within 3.49\%. The major source of the error is the differences in runout distance computed by GNS and MPM. $d_{\phi_{target}}$ and $d_{\phi_{target}}^{GNS}$ represents the runout computed by MPM and GNS at $\phi_{target}$, respectively. Although GNS and MPM use the same friction angle, the GNS includes a small amount of error (1.12\% to 5.58\%) since GNS is a learned surrogate for MPM.

\begin{figure}[]
    \centering
    \includegraphics[width=0.7\textwidth]{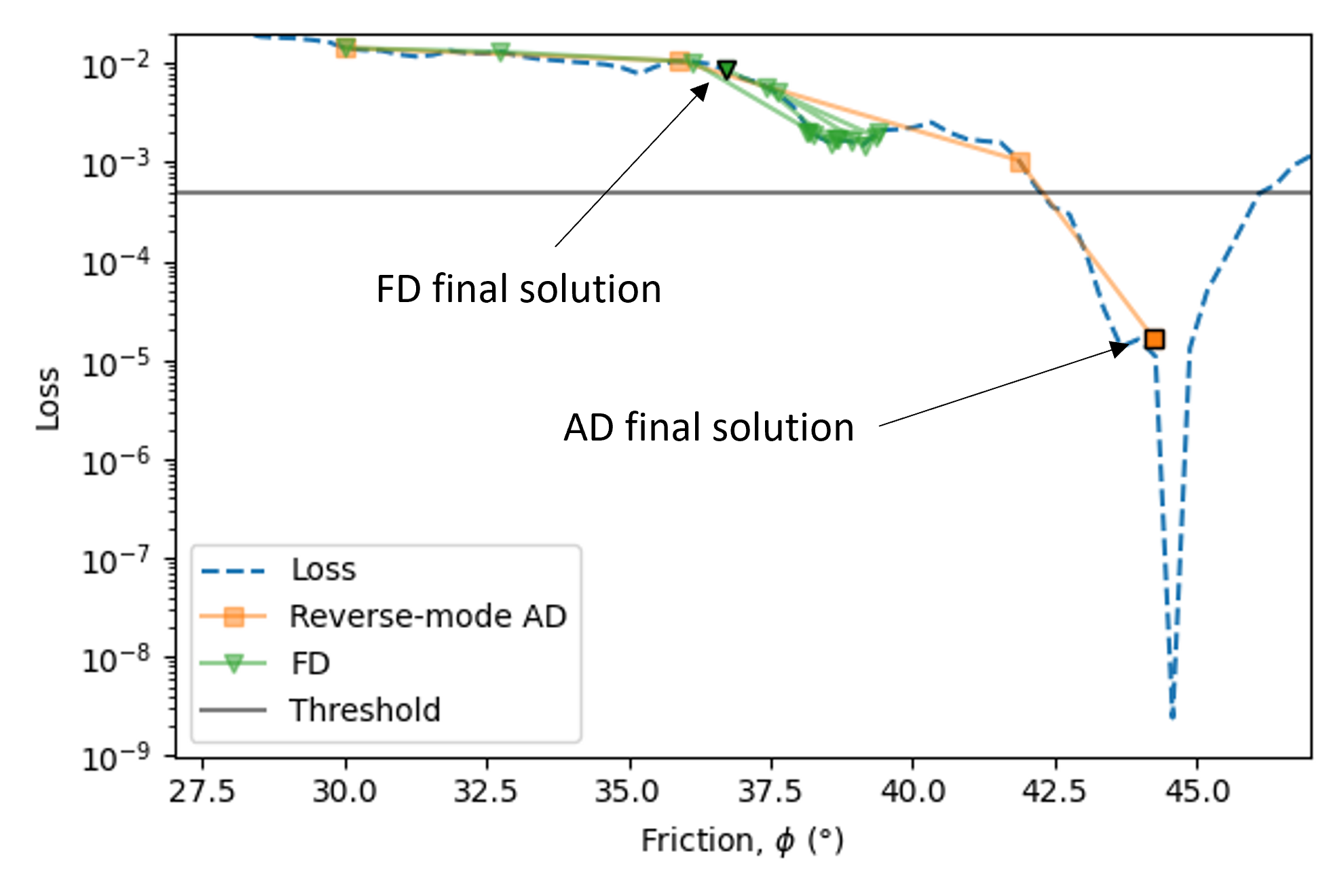}
    \caption{Loss trajectory of reverse mode AD and FD.}
    \label{fig:grad_hist}
\end{figure}

Comparing our AD-based approach to the finite difference (FD) method reveals its stability. \Cref{fig:optimization_phi_history}b shows the optimization history using FD where the gradient is estimated through two forward evaluations at $\phi$ and $\phi + \Delta \phi$. We use $\Delta \phi=0.05$ and set the same learning rate $\eta$ as AD. Despite using the same learning rate as AD, the FD method encounters challenges. The optimization is less stable, and for the tall column with $\phi_{target}=42\degree$, it fails to identify the correct friction angle with the error of about 10\%, which results in $d$ error of 16\%. This error stems from FD's reliance on the chosen $\Delta\phi$ for gradient approximation, unlike AD's precise gradient computations. \Cref{fig:grad_hist} shows the loss history of reverse-mode AD and FD. We observe the loss trajectory of FD being trapped in the flat region between $36\degree$ to  $39\degree$. This is attributed to the diminutive update term $\eta \cdot \nabla_{\phi} J_{\phi}$, a consequence of underestimating gradients due to the selected $\Delta \phi$ value (= 0.05), as opposed to the accurate gradients provided by AD. The first-order forward FD approach has an error of $\mathcal{O}(\Delta \phi)$ in gradient computation.

\begin{table}[]
\centering
\caption{Inverse analysis result for estimating friction angle $\phi$ in granular column collapse. $d_{\phi_{target}}$ is runout distance from MPM with $\phi_{target}$. $d_{\phi_{target}}^{GNS}$ is runout distance from GNS with $\phi_{target}$. $d $ error is percentage error between $d_{\phi_{target}}$ and $d_{\phi_{target}}^{GNS}$. $d_{\phi}^{GNS}$ is runout distance from GNS with estimated $\phi$. $\phi$ error is percentage error between estimated $\phi$ and $\phi_{target}$. $d$ error is percentage error between $d_{\phi}^{GNS}$ and $d_{\phi_{target}}$.}
\label{table:result_phi}
\resizebox{1.0\textwidth}{!}{%
\begin{tabular}{@{}ccccccccccc@{}}
\toprule
\multicolumn{6}{l}{Scenario} & \multicolumn{5}{l}{Optimization} \\ \midrule
\begin{tabular}[c]{@{}c@{}}Column\\ type\end{tabular} & \begin{tabular}[c]{@{}c@{}}$H_0 \times L_0$\\ (m)\end{tabular} & \begin{tabular}[c]{@{}c@{}}$\phi_{target}$\\ ($\degree$)\end{tabular} & \begin{tabular}[c]{@{}c@{}}$d_{\phi_{target}}$\\ (m)\end{tabular} & \begin{tabular}[c]{@{}c@{}}$d_{\phi_{target}}^{GNS}$\\ (m)\end{tabular} & \begin{tabular}[c]{@{}c@{}}$d $ error\\ (\%)\end{tabular} & \begin{tabular}[c]{@{}c@{}}Initial $\phi$\\ ($\degree$)\end{tabular} & \begin{tabular}[c]{@{}c@{}}Estimated $\phi$\\ ($\degree$)\end{tabular} & \begin{tabular}[c]{@{}c@{}}$d_{\phi}^{GNS}$\\ (m)\end{tabular} & \begin{tabular}[c]{@{}c@{}}$\phi$ error\\ (\%)\end{tabular} & \begin{tabular}[c]{@{}c@{}}$d$ error\\ (\%)\end{tabular} \\ \midrule
\multirow{2}{*}{Short} & \multirow{2}{*}{0.2 × 0.4} & 21 & 0.6947 & 0.7154 & 2.98 & 30 & 22.88 & 0.6953 & 8.94 & 0.08 \\ \cmidrule(l){3-11} 
 &  & 42 & 0.6000 & 0.6067 & 1.12 & 30 & 41.31 & 0.6209 & 1.65 & 3.49 \\ \midrule
\multirow{2}{*}{Tall} & \multirow{2}{*}{0.4 × 0.2} & 21 & 0.8318 & 0.8783 & 5.58 & 30 & 22.87 & 0.8401 & 8.90 & 0.99 \\ \cmidrule(l){3-11} 
 &  & 42 & 0.5665 & 0.5942 & 4.89 & 30 & 44.54 & 0.5705 & 6.06 & 0.72 \\ \bottomrule
\end{tabular}
}
\end{table}

In the visual progress of the optimization (see \cref{fig:visual_phi_opt}), the yellow dots represent the final deposit predicted by GNS, and the gray shade represents the final deposit corresponding to $\phi_{target}$ from MPM. For all scenarios, our approach effectively identifies $\phi$ that closely matches $\phi_{target}$, as $d_{\phi}$ converges to $d_{\phi_{target}}$. Notably, the GNS not only accurately predicts the final runout distance but also captures the overall geometry of the granular deposit, although we only use $d_{\phi}$ for the optimization. Unlike typical low-dimensional empirical correlations, GNS offers prediction of full granular dynamics. By simulating the entire granular flow process, GNS enables our optimization to reproduce the runout distance and the detailed granular geometry of the deposit.

\subsection{Multi-parameter inverse} \label{sec:result_multivel}

Real-world inverse problems are complex as they include multiple parameters for optimization. In this section, we evaluate the performance of AD-GNS in solving the multi-parameter boundary condition inverse problem (as outlined in \cref{fig:problem_statement}b). The objective is to determine the initial boundary condition, i.e., $x$-velocities ($\boldsymbol{v}$), of each layer in the multi-layered granular column that produces a target deposit $\boldsymbol{S}_{\boldsymbol{v}_{target}}$. Here, $\boldsymbol{S}_{\boldsymbol{v}}$ is the coordinate value of all material points when the flow stabilizes. Thus, $\boldsymbol{v}$ is our parameter set $\boldsymbol{\Theta}$, and $\boldsymbol{S}_{\boldsymbol{v}}$ is our runout metric to minimize with respect to $\boldsymbol{S}_{\boldsymbol{v}_{target}}$ with an initial $\boldsymbol{v}_{target}$ from MPM simulation.

\begin{figure}[]
    \centering
    \includegraphics[width=0.5\textwidth]{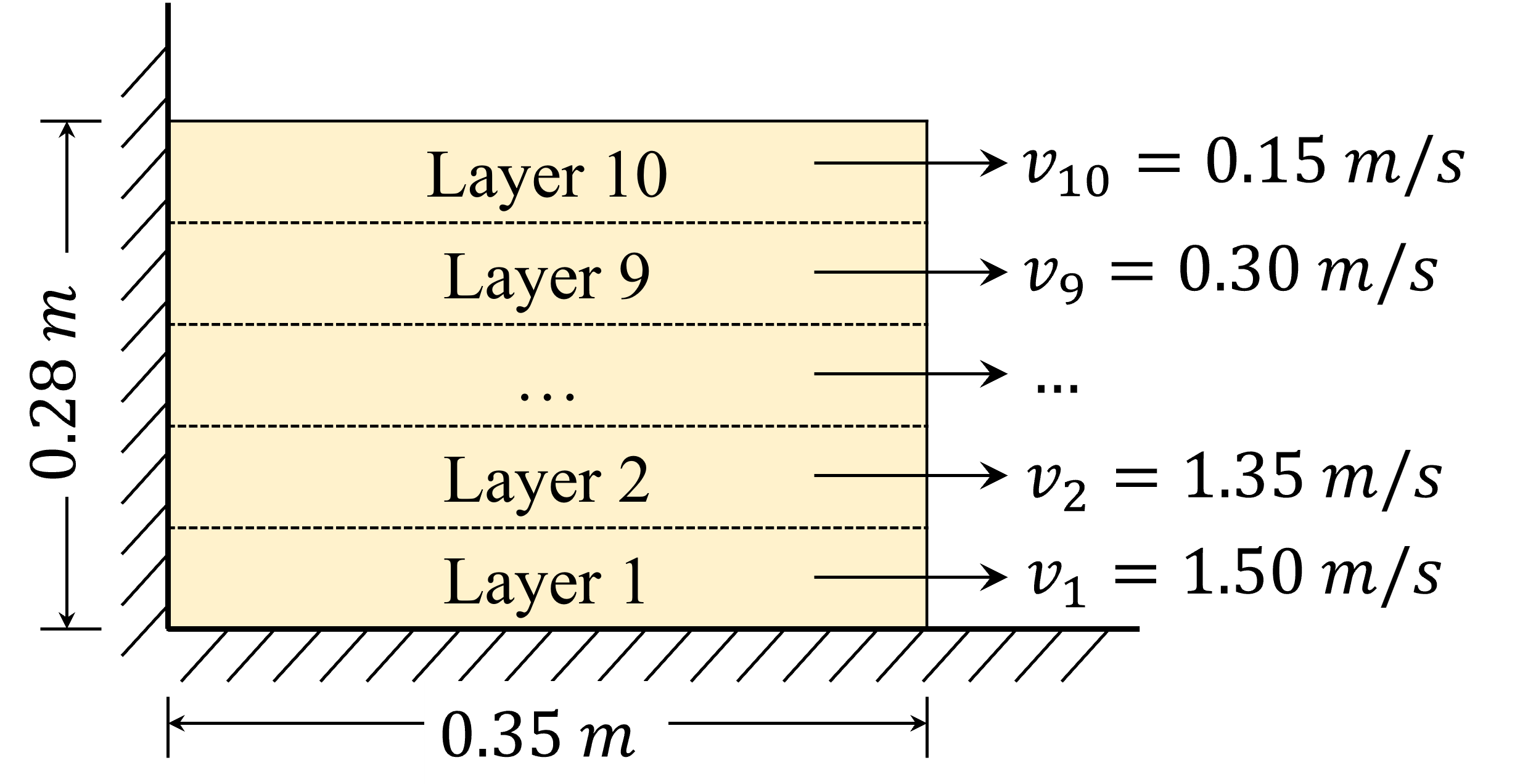}
    \caption{Granular column collapse scenario.}
    \label{fig:layered_column}
\end{figure}

\Cref{fig:layered_column} shows the granular column collapse scenario for the inverse analysis addressed in this section. The length and height of the column is 0.35 × 0.28 m with an aspect ratio of 0.8, including 3,920 particles. It consists of 10 horizontal layers with the same thickness with linearly decreasing initial $x$-velocity from 1.50 m/s at the first layer to 0.15 m/s at the $10^\mathrm{th}$ layer., $\boldsymbol{v}_{target}=[1.50, \ 1.35, \ \ldots, \ 0.30, \ 0.15]$ m/s. The GNS has not encountered this discretized layered velocity condition during training. 

\begin{figure}[]
    \centering
    \includegraphics[width=1.0\textwidth]{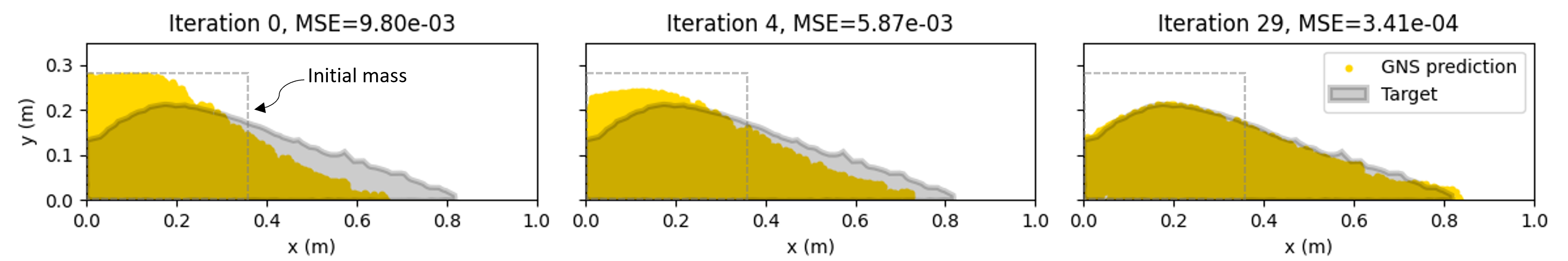}
    \caption{Visualization of GNS prediction during optimization with different $\boldsymbol{v}$. The yellow dots represent the final deposit predicted by GNS ($\boldsymbol{S}_{\boldsymbol{v}}$) and the gray shade represents the target final deposit from MPM with $\boldsymbol{v}_{target}$ ($\boldsymbol{S}_{\boldsymbol{v_{target}}}$)}.
    \label{fig:multivel_deposit_hist}
\end{figure}

\Cref{fig:multivel_deposit_hist} visualizes the GNS prediction of runout during optimization with different $\boldsymbol{v}$.  We update the parameters by updating the ADAM with the learning rate $\eta=0.1$. The loss $J_{\boldsymbol{v}}$ is defined as the mean squared error (MSE) of the coordinate values for materials points between the predicted final deposit $\boldsymbol{S}_{\boldsymbol{v}} = \{ \boldsymbol{p}^i\}_{i=1:N}$ and the target final deposit $\boldsymbol{S}_{\boldsymbol{v_{target}}}  = \{ \boldsymbol{p}_{target}^i\}_{i=1:N}$ from MPM (\cref{eq:multivel loss}). 
Here, $\boldsymbol{p}_{\boldsymbol{v}}^i$ and 
$\boldsymbol{p}^i_{\boldsymbol{v}_{target}}$ denotes the coordinates of $N$ materials points. 

\begin{equation}\label{eq:multivel loss}
J_{\boldsymbol{v}} = \frac{1}{N} \sum_{i=1}^{N} \| \boldsymbol{p}_{\boldsymbol{v}}^i - \boldsymbol{p}^i_{\boldsymbol{v}_{target}} \|^2
\end{equation}

\begin{figure}[]
    \centering
    \includegraphics[width=1.0\textwidth]{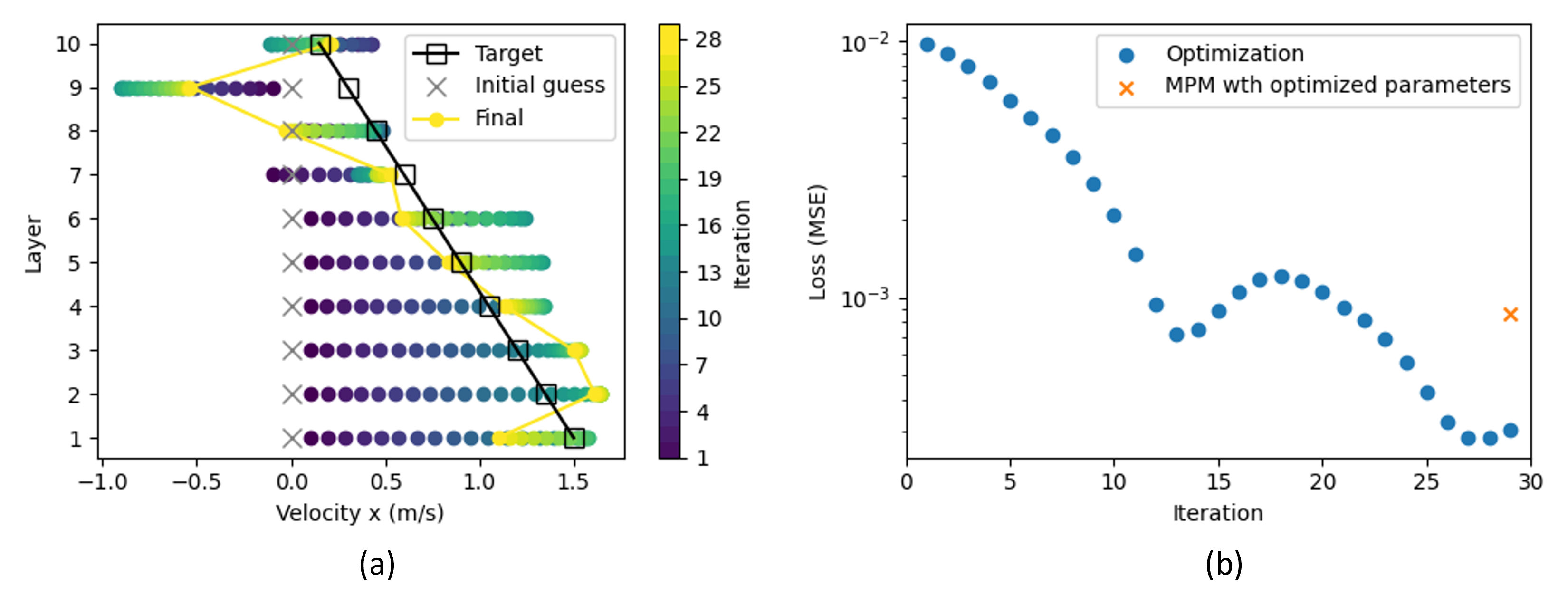}
    \caption{Optimization history for the layer velocities of granular mass: (a) velocities for each layer, (b) loss.}
    \label{fig:multivel_hist}
\end{figure}

\begin{figure}[]
    \centering
    \includegraphics[width=0.5\textwidth]{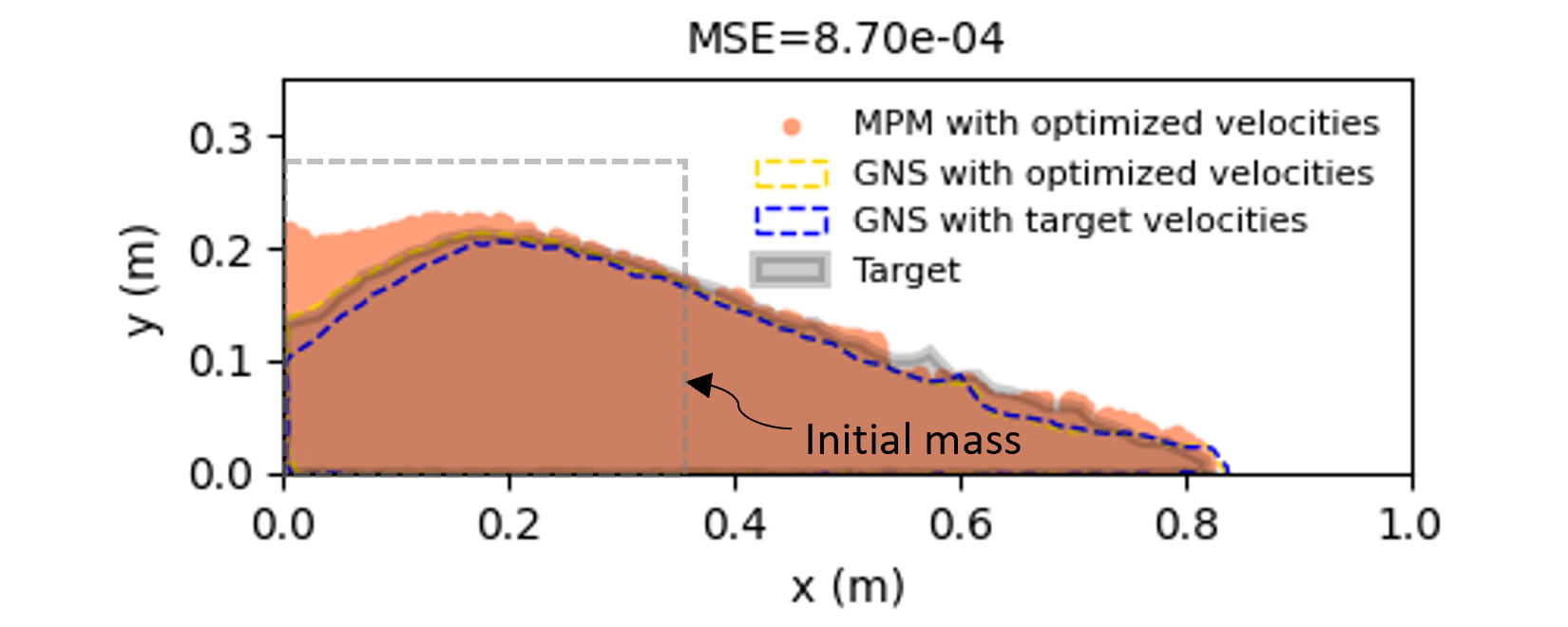}
    \caption{MPM simulation results of the final deposit from the optimized velocities (red dots) and target velocities (gray shade), plotted together with the GNS prediction with target velocities (blue dashed line). The GNS prediction of the final deposit with optimized velocities is plotted with the yellow dashed line, which is the same as the deposit geometry from \cref{fig:multivel_deposit_hist} at the $29^\mathrm{th}$ iteration. The MSE at the top of the figure is for the error between the MPM with optimized velocities and the target.}
    \label{fig:multivel_reconst}
\end{figure}

The gray shade shows the final runout deposit produced from our scenario (\cref{fig:layered_column}) simulated using MPM, which is our target, and the yellow dots are the GNS prediction of the final deposit at the corresponding optimization iterations. The initial velocity guesses for each layer are set to zero, as shown by the cross ($\times$) markers in \cref{fig:multivel_hist}a. As iterations progress, the velocities for each layer gradually align with $\boldsymbol{v}_{target}$, indicated by square markers in \cref{fig:multivel_hist}a, and consequently, the final deposit geometry closely matches the target at iteration 29 with MSE of 3.41e-4, as shown in \cref{fig:multivel_hist}b. \added{We also test the GNS prediction with the target velocities (see \cref{fig:multivel_reconst}). The GNS prediction with the target velocities (blue dashed line in \cref{fig:multivel_reconst}) shows a good agreement with GNS prediction with optimized velocities at iteration 29 (yellow dashed line in \cref{fig:multivel_reconst}) and the target (gray shade in \cref{fig:multivel_reconst}).}

Since GNS is a surrogate model of the forward simulator, it inherently incorporates errors; the inferred velocities might not reproduce the same final deposit geometry in the ground truth simulator, MPM. Hence, verifying if the inferred velocities are still valid in the ground truth simulator when reproducing the target final deposit is necessary. 

To confirm the validity of the inferred velocities, we compare the final deposit with the inferred velocities using MPM with the target final deposit. \Cref{fig:multivel_reconst} shows the MPM simulation results of the final deposit from the inferred velocity at the 29th iteration and target velocity. The gray shade represents the target deposit, and the red dots represent the deposit from MPM using the inferred velocities. The final deposit from the inferred velocity reasonably matches our target with a minor difference at the left boundary with an MSE of 8.70e-4, shown in \cref{fig:multivel_hist}b. The MSE is slightly larger than the optimization error at the $29^\mathrm{th}$ iteration but still significantly smaller than that of the initial iteration, which is about 0.01. The comparison suggests that our method provides reasonable velocity estimations to replicate the target behavior, even though GNS had not been exposed to the discretized layered initial velocity condition during training.

\subsection{Design of baffles to resist debris flow} \label{sec:result_barrier}

We can use AD-GNS to design engineering structures, which involves optimizing the design parameters of structural systems to achieve a specific functional outcome. We demonstrate the use of AD-GNS in the design of the debris-resisting baffles to achieve a target runout distance.

Debris-resisting baffles (\cref{fig:problem_statement}c) are rigid flow-impeding earth structures strategically placed perpendicular to potential landslide paths to interrupt the flow of landslide debris \citep{yang2021}. Their primary function is to reduce the risk and impact of debris flow by decelerating it and dissipating its energy upon impact. The configuration of these baffles is crucial as it directly influences their effectiveness in mitigating debris flow.

Our inverse analysis aims to optimally position the baffles to halt granular flow within a predefined area (as outlined in \cref{fig:problem_statement}c) Specifically, we optimize the $z$-locations of two barriers, denoted as $\boldsymbol{z}$, to ensure that the toe of the granular jet—defined as the centroid $\boldsymbol{c}_{\boldsymbol{z}}$ of the furthest 10\% of downstream particles—stops at a specific target point $\boldsymbol{c}_{\boldsymbol{z}_{target}}$. In this context, $\boldsymbol{z}$ corresponds to our parameter set $\boldsymbol{\Theta}$, and $\boldsymbol{c}_{\boldsymbol{z}}$ represents our measure of runout $\boldsymbol{R}_{\boldsymbol{\Theta}}$ in \cref{fig:inverse_schematic}.    

\begin{figure}[]
    \centering
    \includegraphics[width=0.5\textwidth]{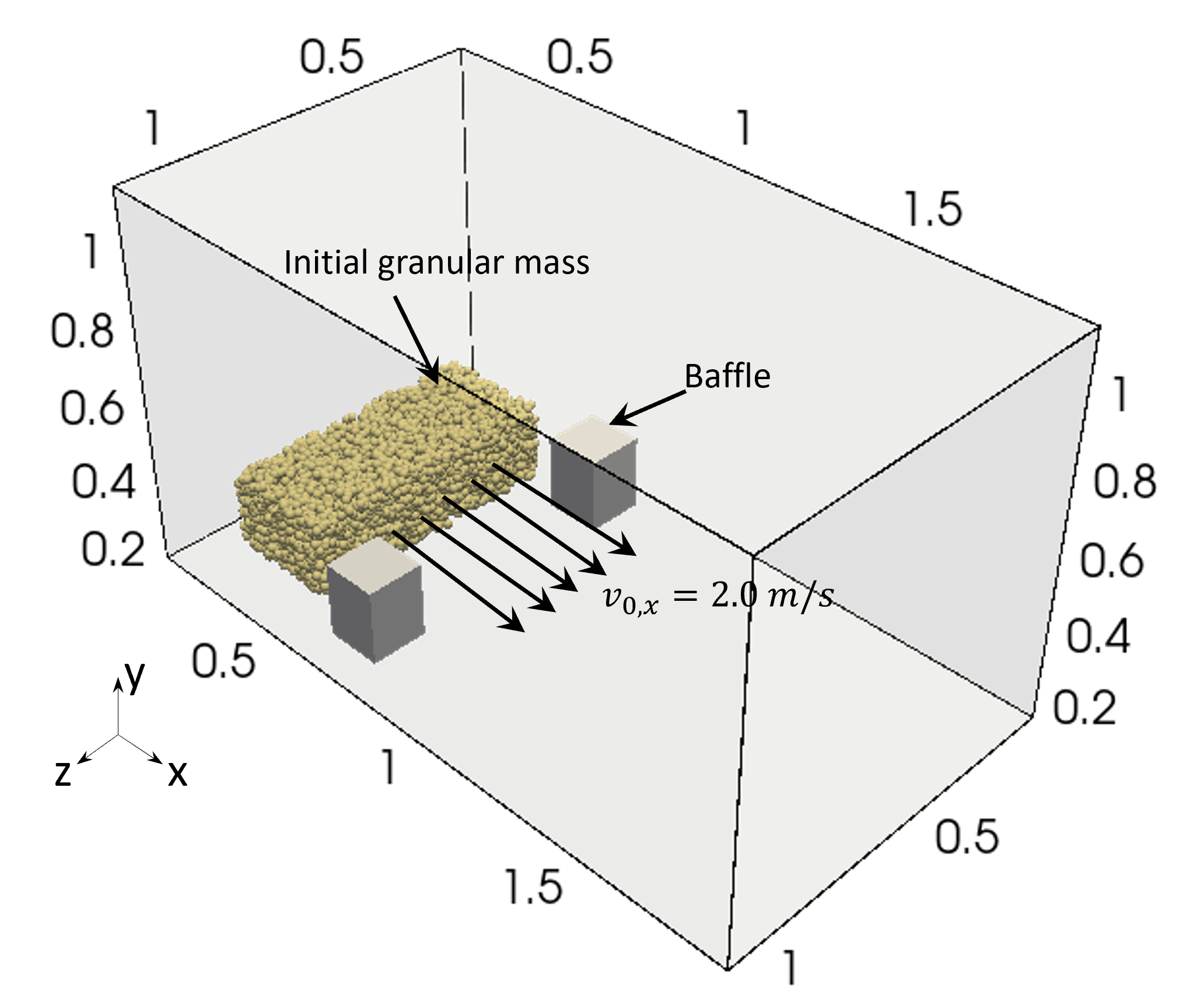}
    \caption{Initial geometry of the granular mass}
    \label{fig:baffle_initial_config}
\end{figure}

The initial geometry of the granular mass (see \cref{fig:baffle_initial_config}) is cuboid-shaped with the size of 0.3 × 0.2 × 0.7 m for $x$, $y$, and $z$ coordinates, and its lower edge is at ($x$=0.2, $y$=0.125, $z$=0.25) m. The granular mass is subjected to a uniform initial $x$-velocity of 2.0 m/s. The size of the baffles is 0.15 × 0.2 × 0.15 m. The total number of particles is 13,371. The simulation domain is 2.0 × 1.0 × 1.0 m. The GNS has never seen this baffle size and the simulation domain during training, which is two times larger than the training domain (see \cref{table:train_data} for comparison). These differences place our inverse solver in a more challenging circumstance. 

\begin{figure}[]
    \centering
    \includegraphics[width=1.0\textwidth]{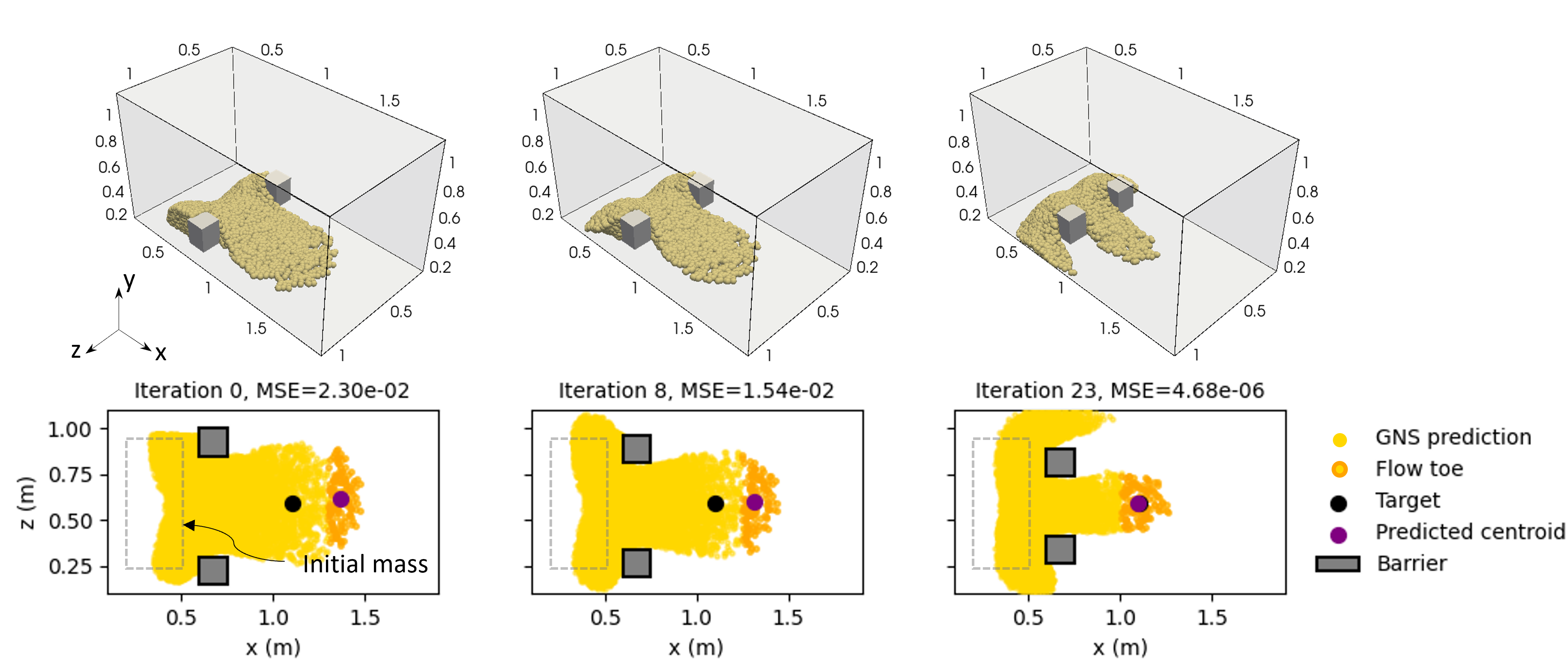}
    \caption{Optimization history for baffle locations}
    \label{fig:barrier_deposit_hist}
\end{figure}

\Cref{fig:barrier_deposit_hist} illustrates the visual progress of the optimization. We use ADAM for the optimization with the learning rate $\eta=0.01$. The loss $J_{\boldsymbol{c}_{\boldsymbol{z}}}$ is defined as the MSE between the coordinates$\boldsymbol{c}_{\boldsymbol{z}}$ and $\boldsymbol{c}_{\boldsymbol{z}_{target}}$, which is the absolute Euclidean distance: 

\begin{equation}\label{eq:design loss}
J_{\boldsymbol{c}_{\boldsymbol{z}}} = \| \boldsymbol{c}_{\boldsymbol{z}} - \boldsymbol{c}_{\boldsymbol{z}_{target}} \|^2
\end{equation}

\begin{figure}[]
    \centering
    \includegraphics[width=0.7\textwidth]{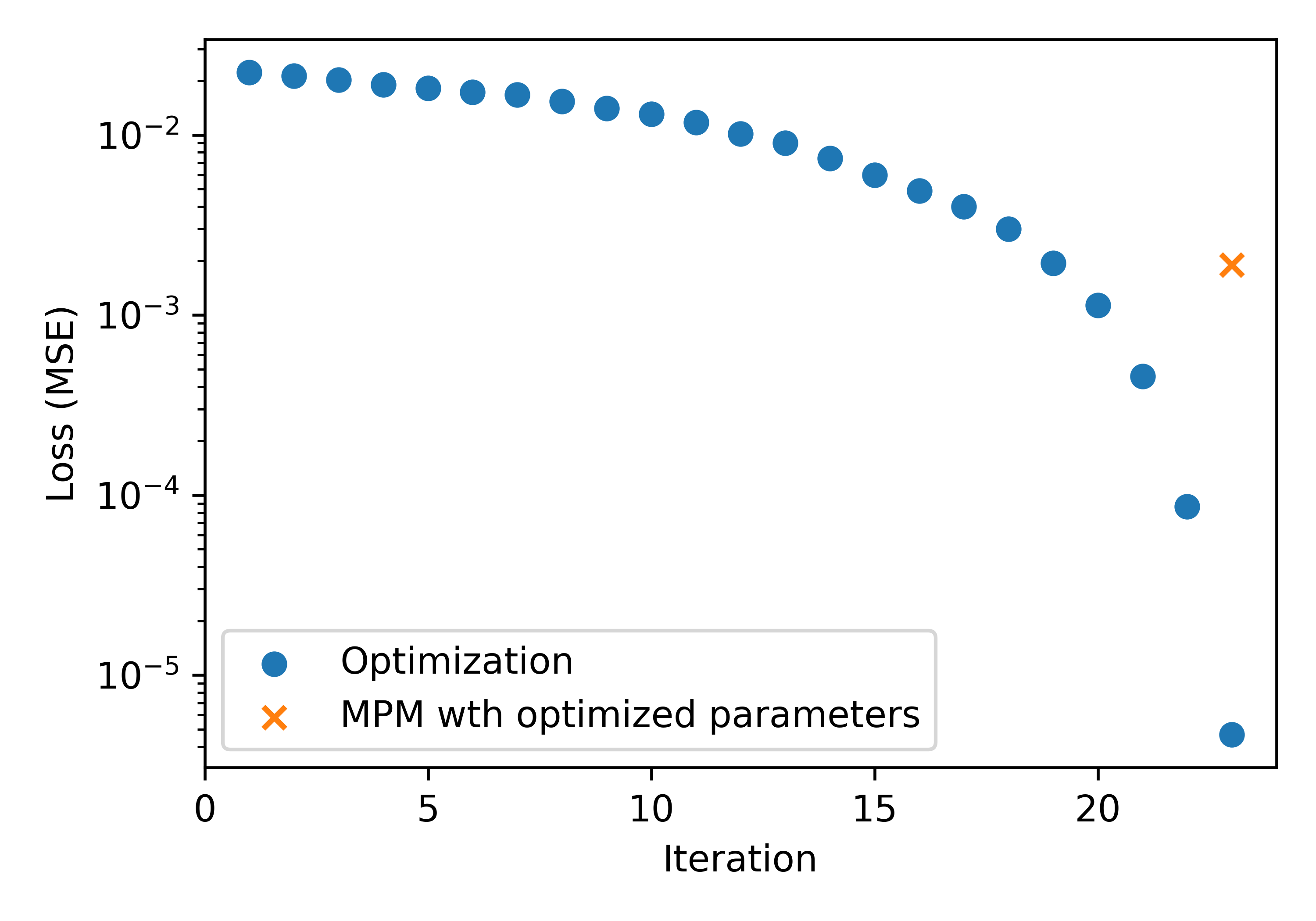}
    \caption{Loss history for the centroid location}
    \label{fig:barrier_loss_hist}
\end{figure}

The gray broken line shows the initial geometry of the granular mass, and the yellow dots represent the final deposit after flow ceases. An orange outline highlights the flow toe, with a purple dot marking its centroid $\boldsymbol{c}_{\boldsymbol{z}}$. The black dot marks the target centroid location $\boldsymbol{c}_{\boldsymbol{z}_{target}}$. We start the initial guess of the barrier center locations at ($x$=0.675 m, $z$=0.225 m) for the lower baffle and at ($x$=0.675 m, $z$=0.925 m) for the upper baffle, as illustrated by the gray boxes in \cref{fig:barrier_deposit_hist}. As iteration proceeds, the centroid of the flow toe converges to the target centroid with the baffles $z$-located at 0.341 m and 0.812 m. As depicted in \cref{fig:barrier_loss_hist}, the loss history converges to a final loss of 4.68e-6 at iteration 23.

\begin{figure}[]
    \centering
    \includegraphics[width=\textwidth]{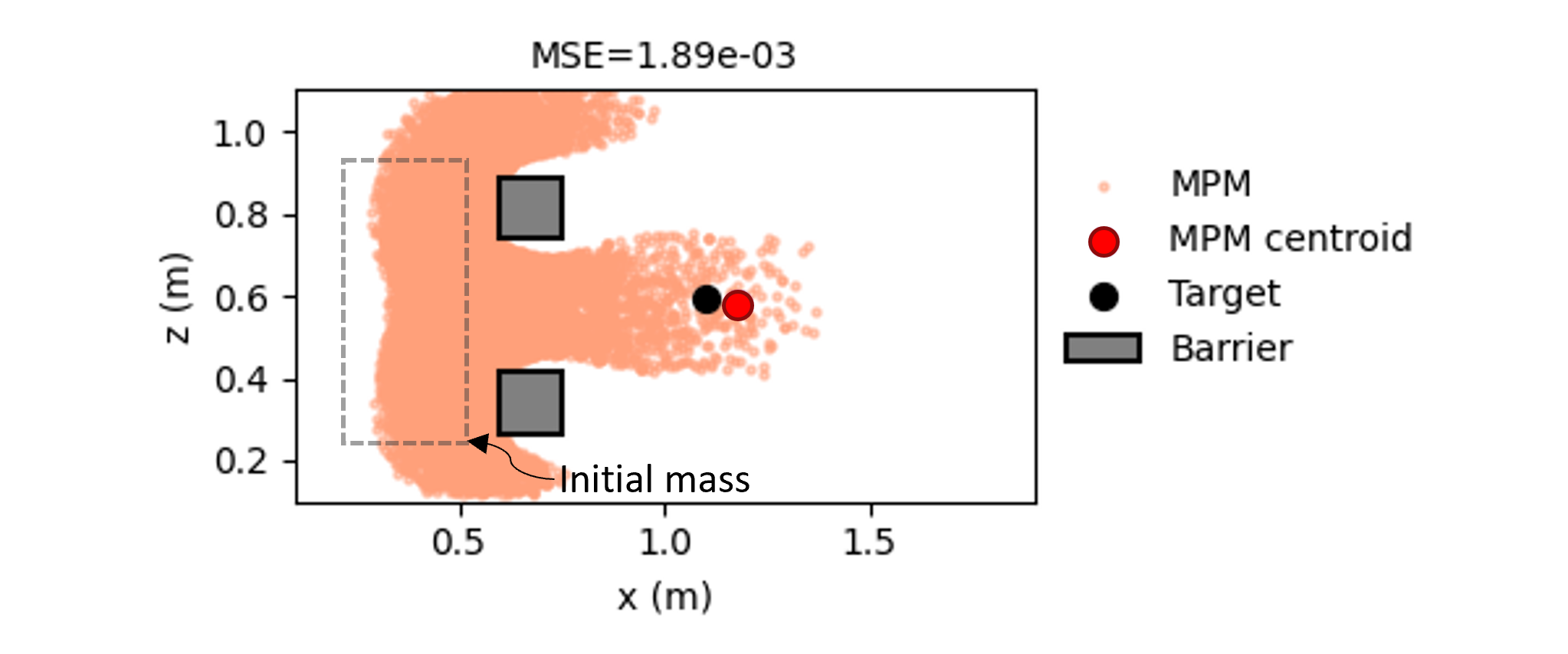}
    \caption{MPM simulation results of final deposit from the inferred baffle locations and its flow toe centroid compared to the target.}
    \label{fig:barrier_deposit_reconst}
\end{figure}

As we discussed in \cref{sec:result_multivel}, we need to validate if the inferred baffle locations are still effective in the ground truth simulator in replicating the desired outcome. \Cref{fig:barrier_deposit_reconst} shows the final deposit and the centroid based on the optimized baffle locations using MPM. Although there is a noticeable difference between the MPM centroid and the target centroid, with the validation loss of 1.89e-3, this deviation is still markedly lower than the loss at the initial iteration (2.30e-2) in \cref{fig:barrier_deposit_hist}. This outcome is promising because we extrapolate beyond the training data.

An essential element in our inverse analysis framework is the implementation of gradient checkpointing on GNS rollout. This technique is highly advantageous, particularly when tracking gradients over long timesteps in 3D scenarios, such as the one described in this section. Without gradient checkpointing, we encountered significant limitations in forward steps due to GPU memory constraints, reaching the 40 GB limit within merely three steps of the rollout for the backpropagation. However, by employing gradient checkpointing and thus storing the sparse results of intermediate steps, we could track gradients effectively for hundreds of steps. This alleviated memory issues and ensured our approach's feasibility for larger and more complex scenarios.

\subsection{Optimization efficiency}

We compare the computation time between our proposed approach and the baseline Finite Difference methods (\cref{table:computation_time}). The results are reported based on the single parameter optimization scenario about the tall column ($a = 2.0$) with $\phi=42\degree$ shown in \cref{fig:visual_phi_opt}d. The baseline method involves executing two forward simulations ($\boldsymbol{X}_0, \ \boldsymbol{X}_1, \ \ldots, \ \boldsymbol{X}_k$), where $k$=374 employing our ground-truth simulator, MPM, at $\phi$ and $\phi+\Delta\phi$ for estimating the gradient using FD. The computation is conducted on 56 Intel Cascade Lake processor cores on the Texas Advanced Computing Center (TACC) Frontera. The mean computation time for the MPM forward simulation takes 8078 s with a standard deviation of 381 s, and it requires the same amount of time to conduct another forward simulation at $\phi+\Delta\phi$ to compute the gradient. Consequently, one optimization iteration requires approximately 8078$\times$2 = 16157 s, which makes the $\phi$ estimation unreasonably time-consuming. Our approach (AD-GNS) does the forward simulation only at $\phi$, and the gradient is computed based on reverse-mode AD. The computation is conducted on RTX with 16 GB memory on TACC Frontera. The forward simulation takes 33 s with the standard deviation of 2.01 s, and the gradient computation using backpropagation takes 74 s with the standard deviation of 1.36 s. Consequently, one optimization iteration requires approximately 107 s on average, outperforming our baseline case by 151 times speedup. The optimization result infers accurate $\phi$ with 6.06\% of error as discussed in \cref{table:result_phi}. 

We also evaluate the computation time for FD with GNS (see \cref{table:computation_time}). The gradient computation from FD with GNS requires less time than AD since it only requires one more forward evaluation in this single parameter case. However, the optimization result from FD shows the unreasonable $\phi$ inference due to inaccurate gradient estimation, in contrast to AD, which infers accurate $\phi$.

For more complex inverse problems that involve high-dimensional parameter space, such as the cases in \cref{sec:result_multivel} and \cref{sec:result_barrier}, FD with MPM becomes almost infeasible due to the computation intensity. Although FD with GNS can return the gradient values faster than MPM, the computation time proportionally increases with the number of parameters, and the gradient values are inaccurate. In contrast, our approach (AD with GNS) can still accomplish efficient and accurate gradient computation owing to the reverse-mode AD despite the increase in the number of parameters, resulting in the successful parameter inference as we showed in \cref{sec:result_multivel} or \cref{sec:result_barrier}.

\begin{table}
\centering
\caption{Mean computation times each optimization iteration for different gradient computation methods. Values are reported based on the short column ($a = 2.0$) with $\phi=42\degree$ for 374 timesteps of simulation duration.}
\label{table:computation_time}
\begin{tblr}{
  cells = {c},
  cell{1}{1} = {r=2}{},
  cell{1}{2} = {c=2}{},
  cell{1}{4} = {c=2}{},
  cell{1}{6} = {r=2}{},
  cell{1}{7} = {r=2}{},
  cell{1}{8} = {r=2}{},
  cell{3}{4} = {c=2}{},
  cell{5}{4} = {c=2}{},
  hline{1,6} = {-}{0.08em},
  hline{2} = {2-5}{},
  hline{3} = {-}{},
}
Method  & {Forward\\simulation (s) } &  & {Gradient\\computation (s) } &  & {Optimization\\iteration (s)  } & {Estimated \\$\phi$} & {$\phi$ error\\(\%) }\\
 & Mean & std & Mean & std &  &  & \\
FD with MPM & 8078 & 381 & 2x of forward &  & 16157 & N/A & N/A\\
AD with GNS & 33 & 2.01 & 74 & 1.36 & 107 & 44.54 & 6.06\\
FD with GNS & 33 & 1.03 & 2x of forward &  & 66 & 37.86 & 9.84
\end{tblr}
\end{table}

\section{Limitations}

GNS is an efficient surrogate model for forward simulation. However, generalization beyond training data may not perfectly replicate the high fidelity ground truth behaviors, resulting in some inevitable error, as shown in \cref{fig:multivel_reconst,fig:barrier_deposit_reconst}. To mitigate this, expanding the training dataset to encompass a more comprehensive array of scenarios could help reduce the prediction error.

\begin{figure}[]
    \centering
    \includegraphics[width=0.7\textwidth]{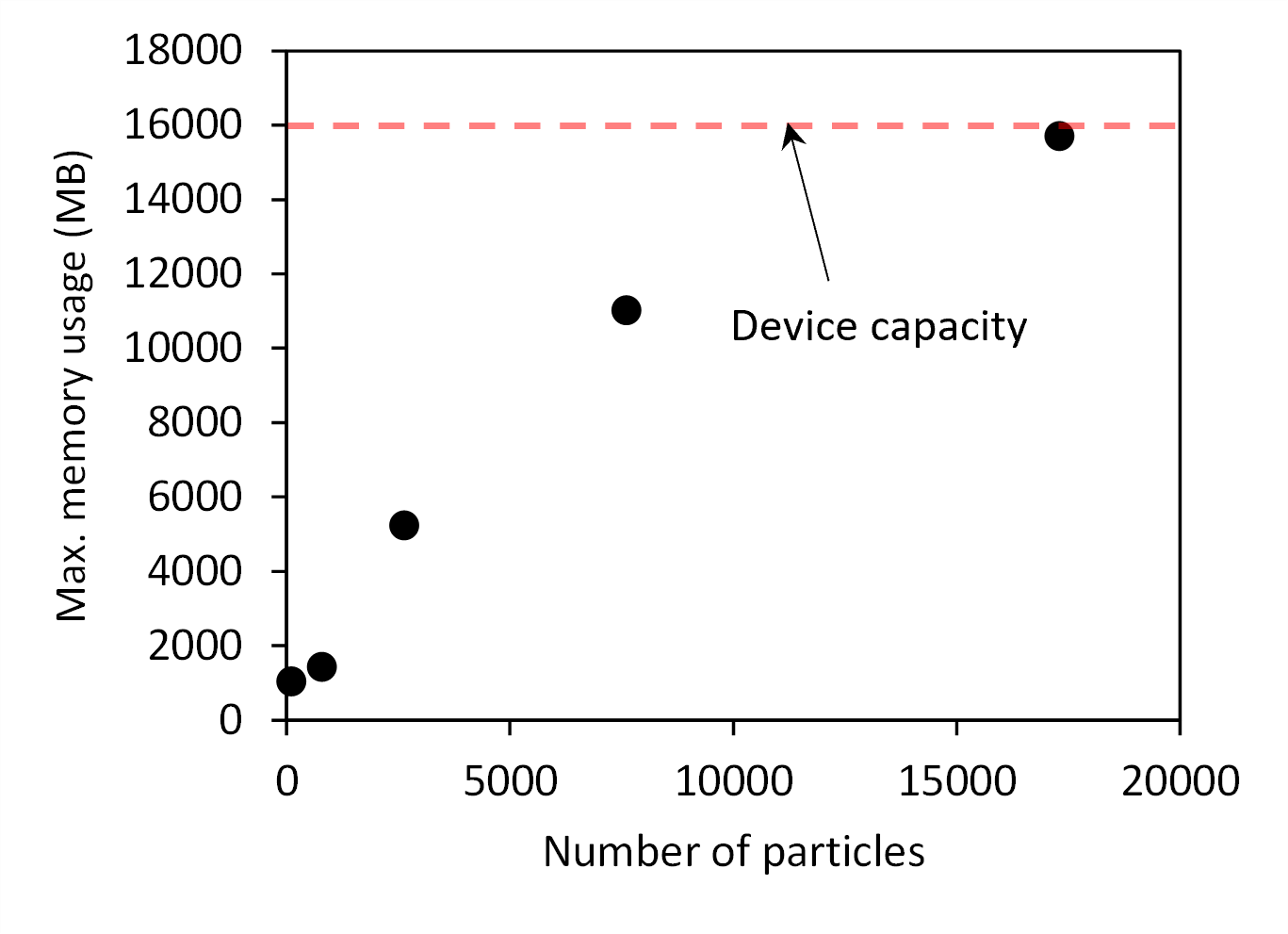}
    \caption{Peak memory usage during backpropagation across up to 400 forward computation steps in relation to an increasing number of particles.}
    \label{fig:checkpointing}
\end{figure}

While gradient checkpointing offers a workaround for the high memory demands associated with reverse-mode AD in GNN, tracking the gradient over the GNN with large graphs is still expensive. We evaluated peak memory usage during backpropagation across up to 400 forward computation steps with increasing particles, as depicted in \cref{fig:checkpointing}. At around 20K particles, the backpropagation process exceeds the memory capacity of our device (16 GB, RTX node at Frontera, TACC), failing. Introducing checkpoints more frequently can reduce memory consumption; however, this strategy requires additional computations for the forward pass, leading to a trade-off between memory efficiency and computational overhead. Another strategy to alleviate memory constraints involves partitioning the graph \citep{ccatalyurek2023graph_partition} and employing multi-GPU processing. Our GNS implementation \citep{kumar2022gns} supports efficient multi-GPU processing. However, effectively partitioning large graphs and facilitating communication across these partitions in a multi-GPU environment presents challenges and is the subject of active research.

\section{Conclusion}

This study introduces an AD-based efficient framework for solving complex inverse problems in granular flows using graph neural network-based simulators (GNS). By leveraging the computational efficiency, differentiability, and generalization capabilities of GNS, coupled with gradient-based optimization through reverse-mode AD, our approach successfully identifies optimal parameters to achieve desired outcomes in diverse granular flow scenarios. 

We demonstrate the effectiveness of our methodology in solving single-parameter and multi-parameter inverse analysis and design problems in granular flows. The single parameter optimization achieves $<$ 8.94\% error in estimating $\phi$ compared to $\phi_{target}$. In the multi-parameter case, the initial velocity is inferred within 8.70e-4 MSE. For the design problem, our method identifies a reasonable baffle arrangement with 1.89e-3 MSE, which is about an order of magnitude improved than the initial guess (2.30e-2). Despite the test configurations being outside the training distribution, our technique generalizes well, highlighting the flexibility of the methodology. Inverse analysis with AD-GNS shows reasonable parameter estimations over a wide range of inverse problems. Nevertheless, the surrogate nature of GNS means some errors are inevitable compared to a high-fidelity simulator. However, the result from our method provides useful information on parameters for further analyses using high-fidelity solvers.

The proposed AD-GNS framework solves inverse problems efficiently on a single GPU, achieving 151x speedups compared to forward high-fidelity simulators with finite difference gradients. Furthermore, the integration of gradient checkpointing enables scaling to complex 3D dynamics over hundreds of timesteps that would otherwise be infeasible due to the excessive memory consumption for backpropagation.

Overall, this study highlights the prospect of data-driven differentiable surrogates in inverse modeling of granular flow hazards. 

\section{Data and code availability}
The code for GNS~\citep{kumar2022gns} and inverse problems are available under the MIT license on GitHub (\url{https://github.com/geoelements/gns} and \url{https://github.com/geoelements/gns-inverse-examples}). The training dataset, trained models, and the data used for the inverse analysis are published under CC-By license on DesignSafe Data Depot~\citep{designsafe-choi-gns, choi2024inverse_data}.

\section{Acknowledgment}
This material is based upon work supported by the National Science Foundation under Grant No.\#2103937 and \#2229702. Any opinions, findings, conclusions, or recommendations expressed in this material are those of the author(s) and do not necessarily reflect the views of the National Science Foundation.

The authors acknowledge the Texas Advanced Computing Center (TACC) at The University of Texas at Austin for providing Frontera and Lonestar6 HPC resources to support GNS training (\url{https://www.tacc.utexas.edu}).




\bibliographystyle{elsarticle-harv} 
\bibliography{main}





\clearpage
\appendix
\setcounter{figure}{0}
\section{Performance of GNS}\label{sec:appendix-performance}
As a benchmark test for Flow2D, we show the performance of GNS using the granular column collapse. Granular column collapse \citep{Lajeunesse2005, utili2015, Lube2005, kumar2015thesis} is an experiment to study the dynamics of granular flows in a controlled setting. A granular column of initial height $H_0$ and length $L_0$ is placed on a flat surface and allowed to collapse under gravity. The dynamics of granular column collapse is majorly governed by the initial aspect ratio of the column ($a = H_0 / L_0$). The previous study \cite{choi2023_gns-column} proves that GNS can accurately predict different flow dynamics of granular columns with various aspect ratios beyond which the data the model is trained. In this paper, we briefly show the prediction performance of the GNS used in this study on a short column ($a=0.5$) with different friction angles ($\phi=21 \degree$ and $42 \degree$), whose aspect ratio and friction angles are not seen during the training.

\begin{figure}[!htbp]
     \centering
     \begin{subfigure}[b]{0.49\textwidth}
         \centering
         \includegraphics[width=\textwidth]{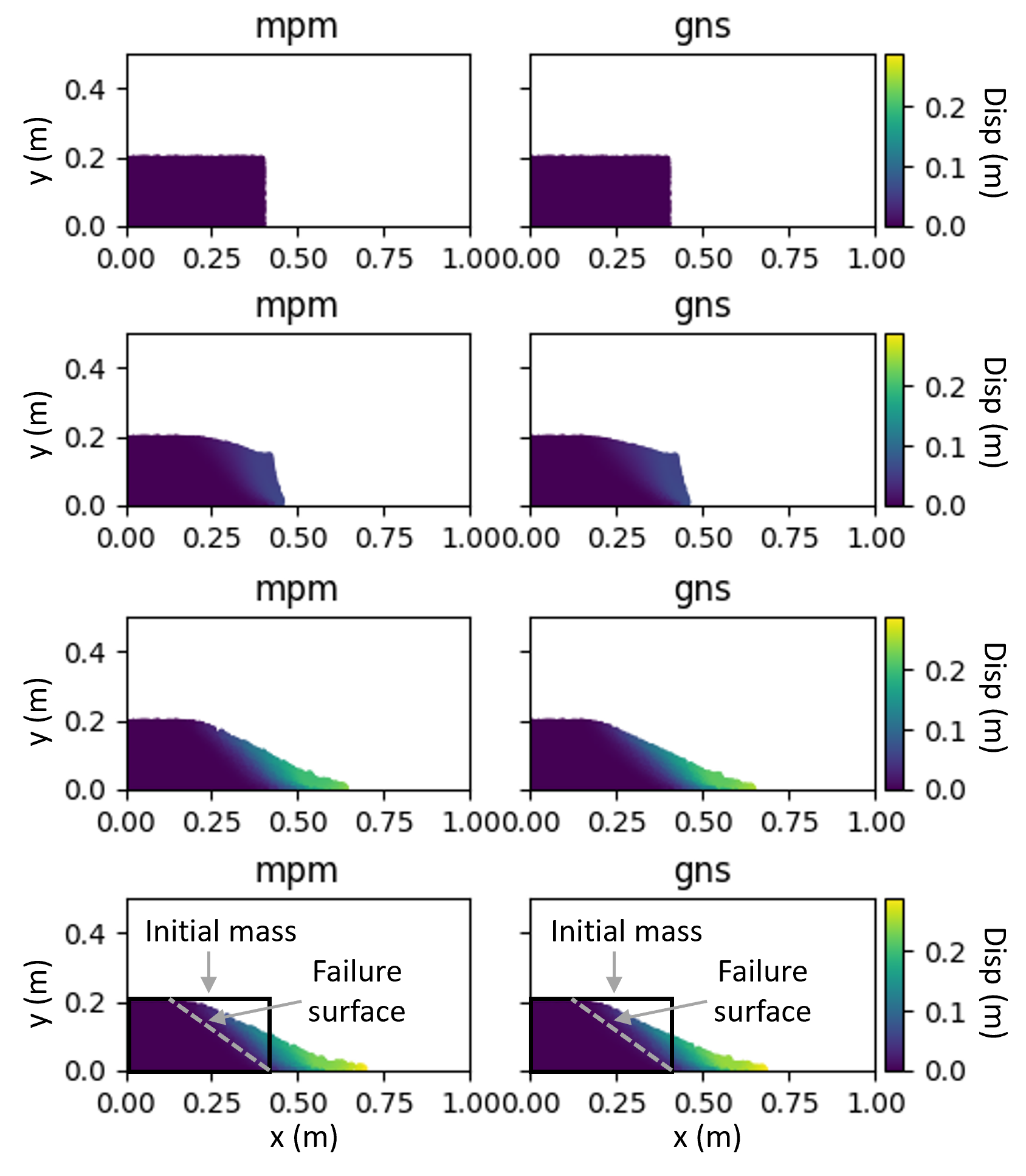}
        \caption{$\phi=21 \degree$}
         \label{fig:progress_short_phi21}
     \end{subfigure}
     \hfill
     \begin{subfigure}[b]{0.49\textwidth}
         \centering
         \includegraphics[width=\textwidth]{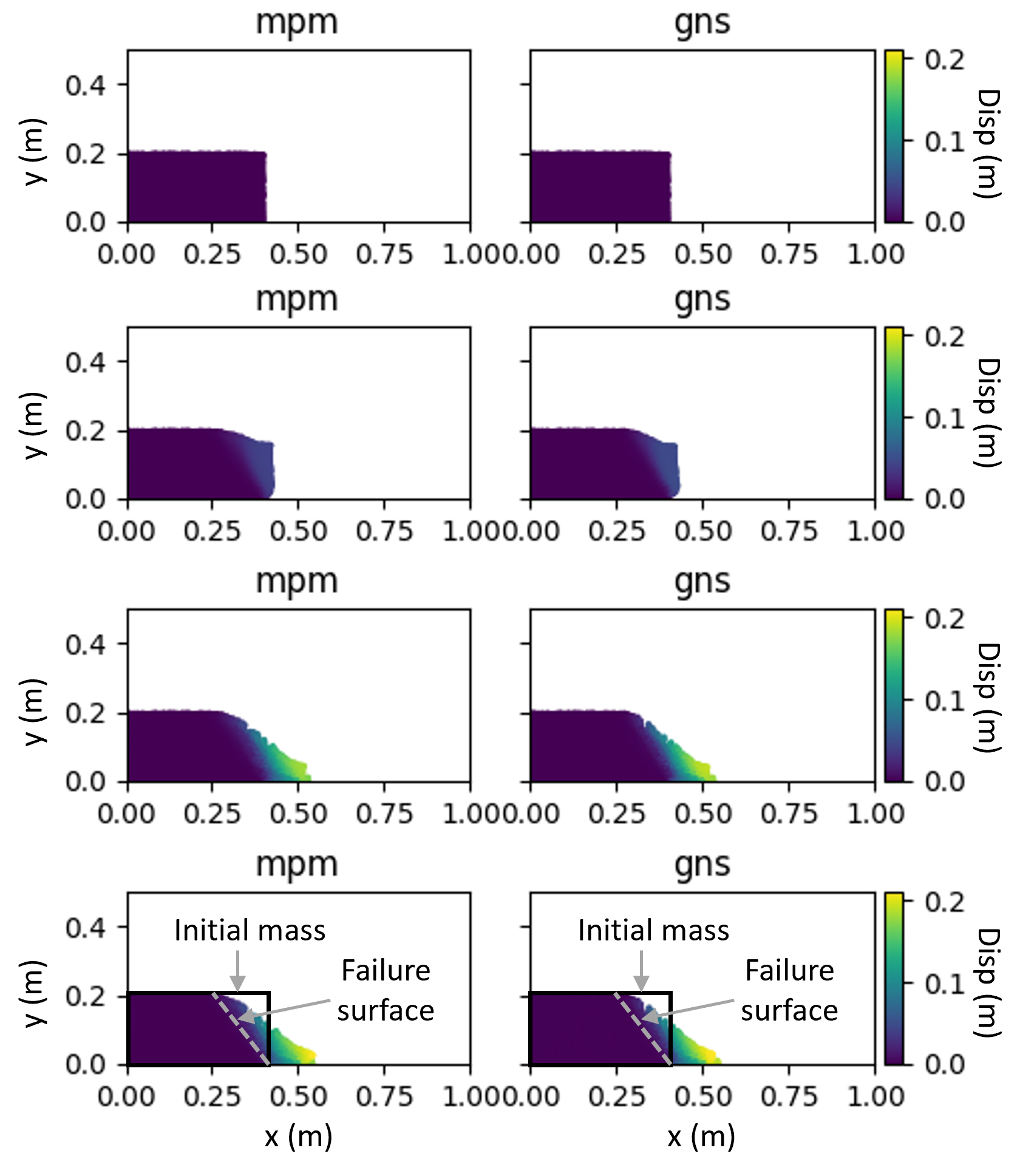}
         \caption{$\phi=42\degree$}
         \label{fig:progress_short_phi42}
     \end{subfigure}
     \caption{Evolution of material point flow with normalized time for GNS and MPM for the short column with a = 0.5: (a) $\phi=21 \degree$, (b) $\phi=42 \degree$.  The color represents the magnitude of the displacement. Each row corresponds to a column before the flow initiation,  $t/\tau_c=1.0$, $t/\tau_c=2.5$, and the final deposit at the last timestep.}
    \label{fig:short_column}
 \end{figure}

\Cref{fig:short_column} shows the evolution of granular flow for the short column ($a=0.5$) with normalized time ($t/\tau_c$) simulated by GNS and MPM. Here, we assume the MPM result is the ground truth. $t$ is physical time, and $\tau_c$ is the critical time, defined as the time required for the flow to mobilize fully. $\tau_c$ is defined as $\sqrt{H_0/g}$, where $g$ is the gravitational acceleration. \Cref{fig:progress_short_phi21} and \cref{fig:progress_short_phi42} is the result with $\phi=21\degree$ and $\phi=42\degree$. Each row of the figure corresponds to flow (1) at the time before initiation of collapse, (2) $t/\tau_c = 1.0$, (3) $2.5$, and (4) at the last timestep ($k=380$) when the flow reaches static equilibrium and forms the final deposit. 

Generally, the collapse shows three stages for both friction angles. First, the flow is mobilized by the failure of the flank and reaches full mobilization around $t/\tau_c=\ 1.0$. Next, the majority of the runout occurs until $t/\tau_c=2.5$. Beyond $t/\tau_c= 2.5$, the spreading decelerates due to the basal friction and finally stops. \Cref{fig:short_column} shows GNS successfully captures this overall flow progress stages for the different friction angles. 

Although the overall progress of the flow is similar, the detailed behaviors diverge depending on the friction angles. When $\phi$ is small (\cref{fig:short_column}a), a larger amount of flow mass is mobilized along the flank of the column above the failure surface until $t/\tau_c=1.0$ compared to the column with larger $\phi$ (\cref{fig:short_column}b). The greater mobilization of the flank failure in small $\phi$ leads to a longer runout until the end of the flow than that of large $\phi$.  At the end of the flow, a greater amount of static soil mass is observed below the failure surface for larger $\phi$ with a truncated conical shape. In addition, a smaller plateau is observed at the top of the final deposit in the case of smaller $\phi$. The visual comparison of the flow profile (\cref{fig:short_column}) shows that GNS well replicates the different granular flow behaviors depending on friction angle with the final runout error of 2.09\% and 1.17\% for each case. 

\begin{figure}[!htbp]
    \centering
    \includegraphics[width=1.0\textwidth]{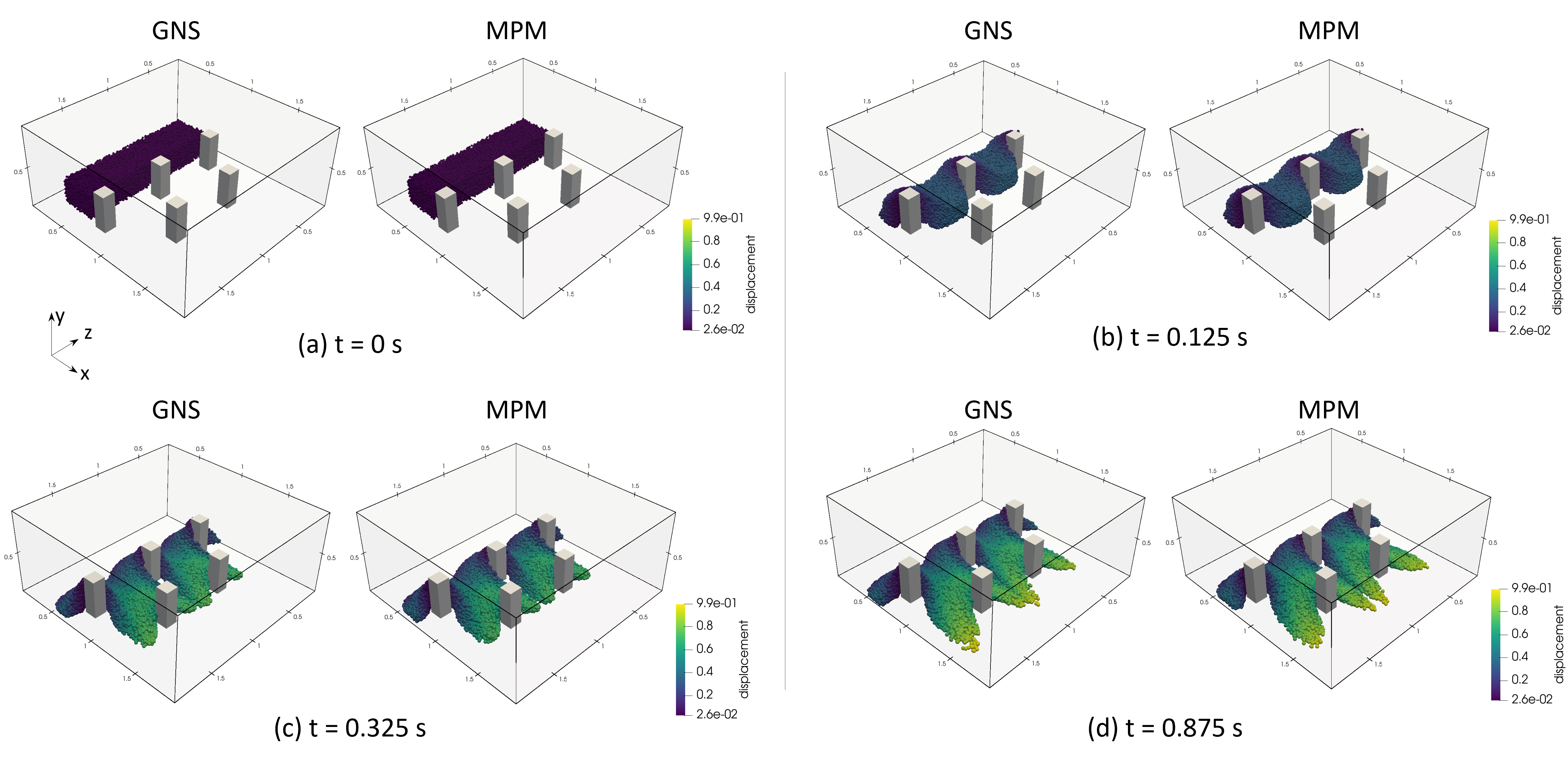}
    \caption{Evolution of flow interacting with baffles for GNS and MPM from initial
condition to the final deposit. The simulation domain is $1.8 \times 0.8 \times 1.8 \ m$, and the initial geometry of the granular mass is $0.35\times0.25\times1.4 \ m$. The barrier size is  $0.15\times0.30\times0.15 \ m$. The center location of the three baffles in the first row is at $(x=0.76, \ z=0.36) \ m$,  $(x=0.76, \ z=1.00) \ m$, and $(x=0.76, \ z=1.64) \ m$, and the two baffles in the second row  are at $(x=1.26, \ z=0.68) \ m$ and $(x=1.26, \ z=1.32) \ m$.}
    \label{fig:progress-barrier}
\end{figure}

Figure \ref{fig:progress-barrier} shows an example of GNS prediction trained on Obstacle3D datasets. The test configuration is outside the training data distribution as summarized in \ref{table:train_data}. Specifically, the simulation domain is four times larger, with 2.2 times more material points than the training data. The maximum length of the initial granular mass in our test is 1.7 m, while that of training data is 0.7 m. Additionally, we test the GNS with five baffles, versus one to three baffles in the training data.

From the initial state to t=0.125 s (\cref{fig:progress-barrier}a-b), the granular debris propagates downstream uniformly and impacts the first baffle row. Upon hitting this row, the baffles obstruct the sides of the flow while the material between the baffles proceeds towards the next row. Concurrently, material deposits and builds up upstream of the baffles, reaching almost the height of the baffle. From t=0.125 s to t=0.325 s (\cref{fig:progress-barrier}b-c), the flow impacts the next baffle row and diverges into four granular jets through the open area with roughly symmetric shapes. Less material deposition forms upstream of the last baffle row than the first. Beyond t=0.325 s (\cref{fig:progress-barrier}c-d), the spreading of the grains decelerates due to basal friction and finally reaches static equilibrium around t=0.875 s. At this stage, the runout error between GNS and MPM is 0.87\%. The GNS rollout successfully replicates the overall kinematics, including complex baffle interactions.

\begin{figure}[!htbp]
     \centering
     \begin{subfigure}[b]{0.49\textwidth}
         \centering
         \includegraphics[width=\textwidth]{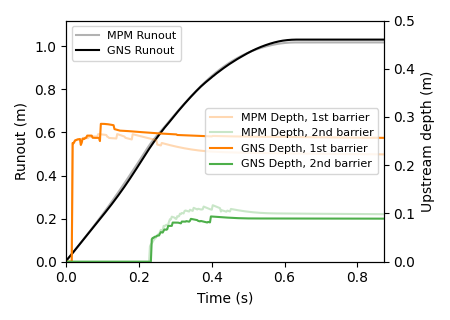}
         \caption{}
         \label{fig:baffle5-runout}
     \end{subfigure}
     \hfill
     \begin{subfigure}[b]{0.49\textwidth}
         \centering
         \includegraphics[width=\textwidth]{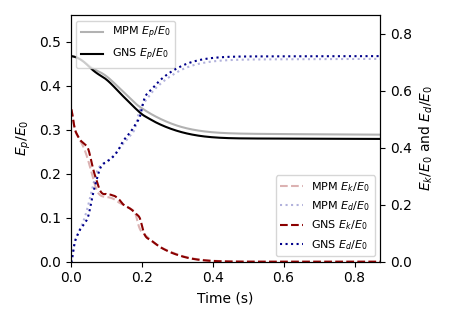}
         \caption{}
         \label{fig:baffle5-energy}
     \end{subfigure}
     \caption{Quantitative comparison between MPM and GNS: (a) Runout and upstream depth evolution with time, (b) potential, kinetic and dissipation energy ($E_p$, $E_k$, and $E_d$) evolution with time normalized by $E_0$, where $E_0$ is total energy at $t=0$.}
    \label{fig:baffle5-quantitative_comparison}
 \end{figure}

To quantitatively compare the results from the GNS and MPM, the evolution of the runout distance is measured (\cref{fig:baffle5-quantitative_comparison}a). Here, runout is defined as the flow's maximum travel distance compared to the initial granular mass geometry. As flow initiates, the runout rapidly increases until 0.325. Subsequently, the runout gradually decelerates and eventually stops due to the interaction with the baffles and frictional dissipation. GNS accurately predicts the runout evolution simulated by MPM despite the simulation including more barriers and a larger domain than the training data.

The upstream depth evolution of the granular mass deposition behind the baffles is also measured (\cref{fig:baffle5-quantitative_comparison}a). The upstream flow depth refers to the depth measured upstream of the baffle array when the flow impacts the baffles. As the runout proceeds, the flow faces the first and then a second row of baffles. Upon impact with the first baffles, the upstream depth spikes close to the baffle height (0.3 m) due to the deposition of the granular mass. It then decreases slightly over time as the flow around the baffles reaches equilibrium. When the flow hits the next baffle row, the upstream depth surges again and gradually decreases. GNS accurately replicates the overall upstream depth change trend with minor errors.

\Cref{fig:baffle5-quantitative_comparison}b shows the energy evolution. The initial granular flow shows high kinetic and potential energy. Subsequent granular flow and its blockage due to the baffles cause the kinetic and potential energy drop. GNS precisely captures these energy evolution trends. The results from \cref{fig:baffle5-quantitative_comparison} suggest that GNS successfully learns the granular flow dynamics with obstacle interactions.

\begin{figure}[!htbp]
    \centering
    \includegraphics[width=0.7\textwidth]{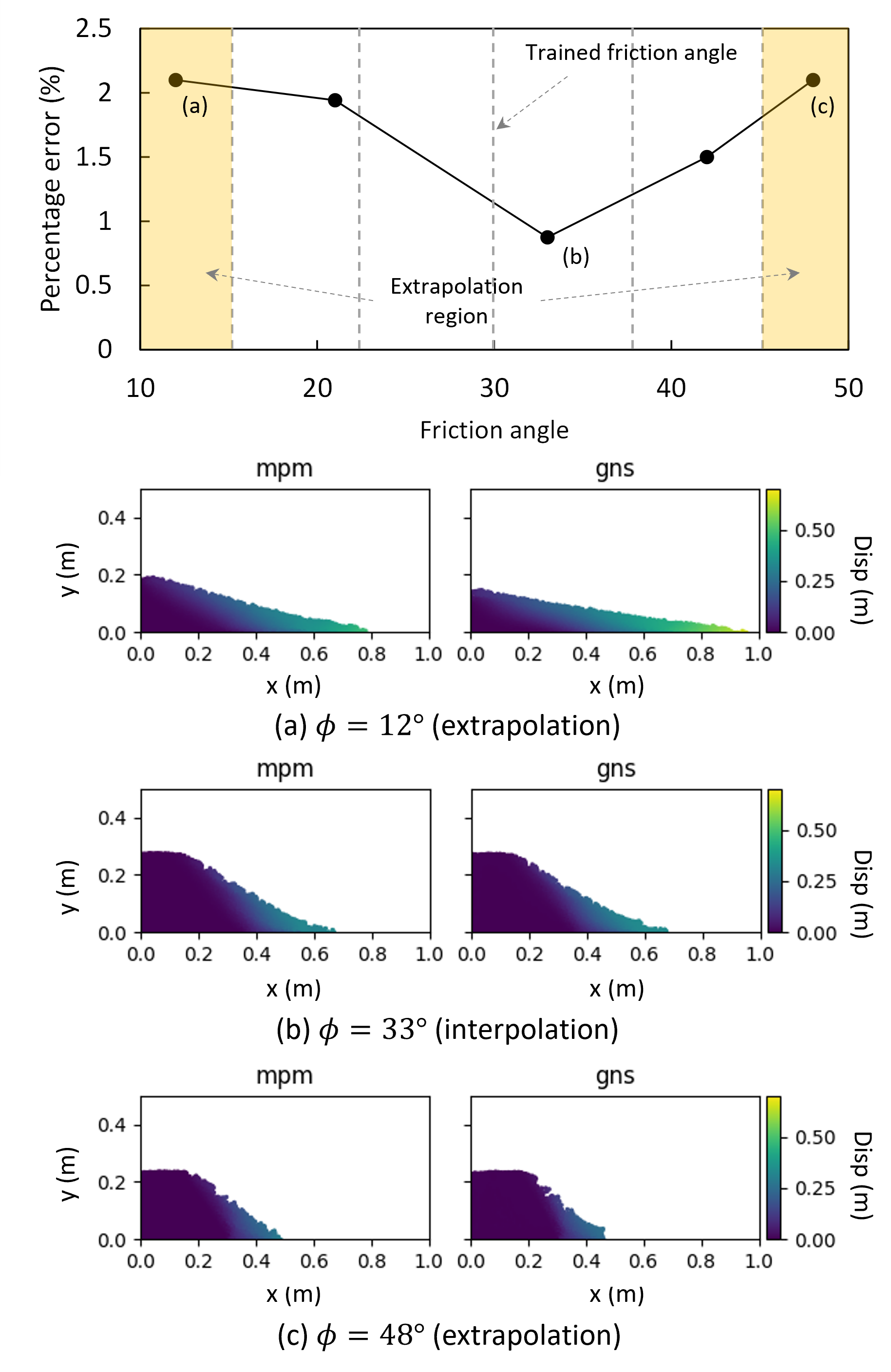}
    \caption{Normalized runout error between MPM and GNS evaluated at different friction angles $\phi$. The final deposit at points (a), (b), and (c) is shown below in the plot for the normalized runout error.}
    \label{fig:extrapolation}
\end{figure}

As explained in \cref{table:train_data}, the dataset for training the 2D GNS (Flow2D) only contains the granular flow trajectories with the friction angles $\phi=15, \ 22.5, \ 30, \ 37.5, \ 45 \degree$. Here, we show the performance of GNS on the friction angles not observed during training. The \cref{fig:extrapolation} shows the normalized runout prediction error (\%) for the granular column with $a$=0.8. The gray dashed lines represent the training $\phi$. The test is conducted with $\phi=12, \ 21, \ 33, \ 42, \ 48 \degree$. In particular, $\phi=12\degree$ and $48\degree$ lies on the extrapolation region as highlighted in \cref{fig:extrapolation}. Although the extrapolation shows a slightly higher error than the predictions on the interpolation region ($\phi=21, \ 33, \ 42 \degree$), the GNS produces good runout predictions for the friction angles not seen during the training.

\section{Training data}\label{sec:appendix-train_data}
\setcounter{figure}{0}

\begin{figure}[!htbp]
    \centering
    \includegraphics[width=0.8\textwidth]{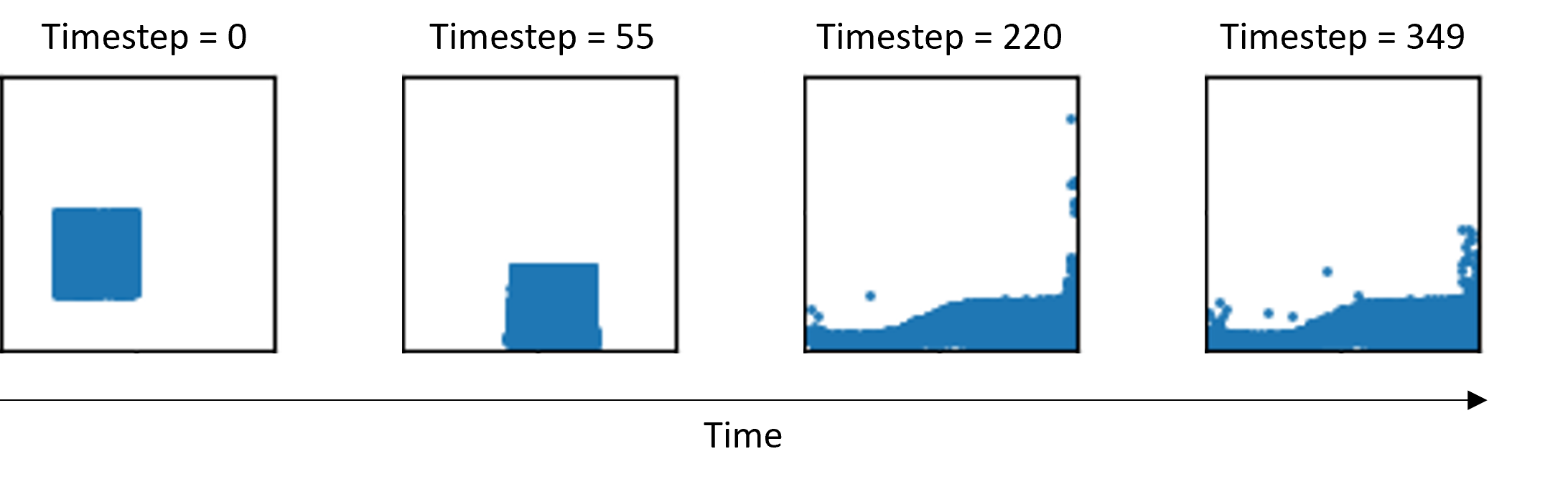}
    \caption{An example of training data for 2D GNS (Flow2D).}
    \label{fig:train_data_2d}
\end{figure}

\begin{figure}[!htbp]
    \centering
    \includegraphics[width=0.8\textwidth]{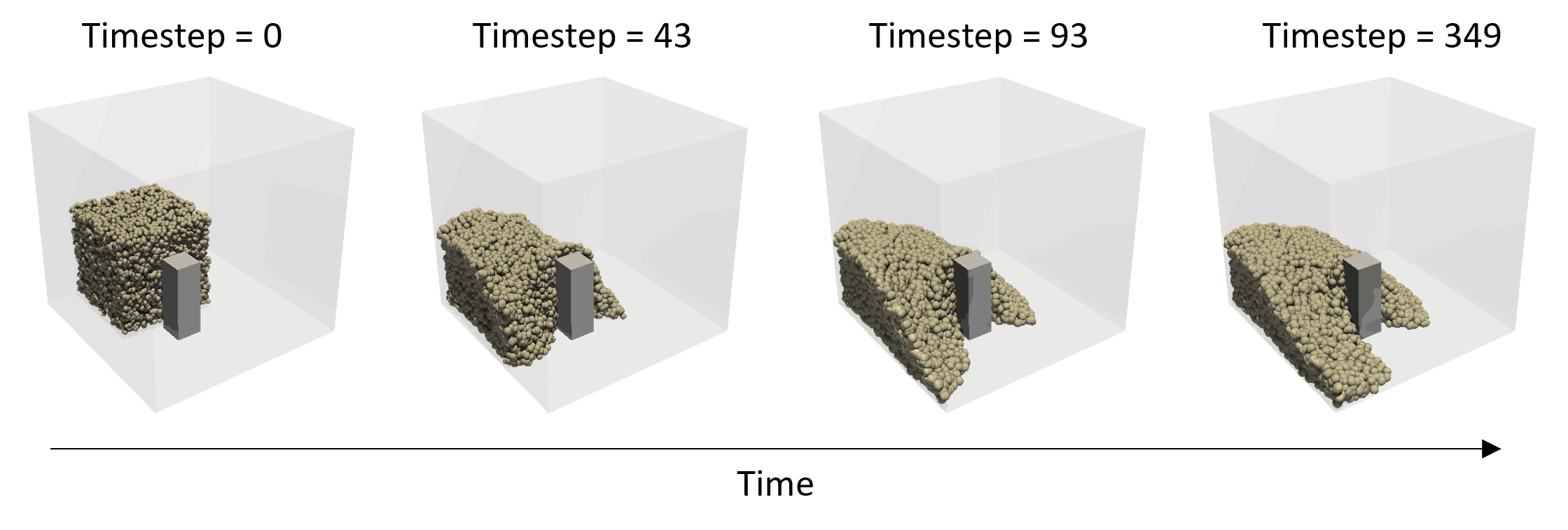}
    \caption{An example of training data for 3D GNS (Obstacle3D).}
    \label{fig:train_data_3d}
\end{figure}

\end{document}